\RequirePackage{lineno}
\documentclass[aip,superscriptaddress,showpacs]{revtex4}

\usepackage[T1]{fontenc}
\usepackage{color,graphicx}
\usepackage{times}
\usepackage[colorlinks=true]{hyperref}
\usepackage{amsmath,amsthm,amssymb}
\usepackage{setspace}
\usepackage{cleveref}
\usepackage{palatino}
\usepackage{pgfplotstable}
\usepackage{pgfplots}
\usepackage{mathdots}
\usepackage{natbib}
\usepackage{enumerate}
\usepackage{doi}
\usepackage{tikz}
\usepackage{subfigure}
\usepackage{multirow}
\usepackage{hhline}

\newcolumntype{C}[1]{>{\centering\arraybackslash}p{#1}}

\usetikzlibrary{backgrounds,fit,decorations.pathreplacing,calc,arrows,automata,shapes}

\newcommand\ket[1]{\ensuremath{|#1\rangle}}
\newcommand\bra[1]{\ensuremath{\langle#1|}}

\begin{document}

\ifdefined\lineno
	\linenumbers
\else
\fi

%% End-Of-Header

\title{Classical Simulation of Intermediate-Size Quantum Circuits}

\author{Jianxin Chen}%
\email{liangjian.cjx@alibaba-inc.com}
\affiliation{Aliyun Quantum Laboratory, Alibaba Group, Hangzhou, Zhejiang 311121, China}%

\author{Fang Zhang}%
\email{fang.z@alibaba-inc.com}
\affiliation{Aliyun Quantum Laboratory, Alibaba Group, Bellevue, WA 98004, USA}%
\affiliation{Department of Electrical Engineering and Computer Science, University of Michigan, Ann Arbor, MI 48109, USA}%

\author{Cupjin Huang}%
\affiliation{Aliyun Quantum Laboratory, Alibaba Group, Bellevue, WA 98004, USA}%
\affiliation{Department of Electrical Engineering and Computer Science, University of Michigan, Ann Arbor, MI 48109, USA}%

\author{Michael Newman}%
\affiliation{Aliyun Quantum Laboratory, Alibaba Group, Bellevue, WA 98004, USA}%
\affiliation{Department of Mathematics, University of Michigan, Ann Arbor, MI 48109, USA}%

\author{Yaoyun Shi}%
\affiliation{Aliyun Quantum Laboratory, Alibaba Group, Bellevue, WA 98004, USA}%

\begin{abstract}
We introduce a distributed classical simulation algorithm for general quantum circuits, and present numerical results for calculating the output probabilities of universal random circuits. We find that we can simulate more qubits to greater depth than previously reported using the cluster supported by the Data Infrastructure and Search Technology Division of the Alibaba Group.  For example, computing a single amplitude of an $8\times 8$ qubit circuit with depth $40$ was previously beyond the reach of supercomputers. Our algorithm can compute this within $2$ minutes using a small portion ($\approx$ 14\% of the nodes) of the cluster. 

Furthermore, by successfully simulating quantum supremacy circuits of size $9\times 9\times 40$, $10\times 10\times 35 $, $11\times 11\times 31$, and $12\times 12\times 27 $,  we give evidence that noisy random circuits with realistic physical parameters may be simulated classically.  This suggests that either harder circuits or error-correction may be vital for achieving quantum supremacy from random circuit sampling.

\end{abstract}

\date{\today}


\maketitle

\section{Introduction}\label{intro}

Classically simulating quantum systems is a relatively old problem \cite{feynman1982simulating}.  However, only recently have nascent quantum computers become competitive in simulating general quantum circuits. Recent announcements of larger systems with reasonable target fidelities \cite{kelly2015state,song201710} are pushing the boundary of what classical simulations can handle.  With this push, a variety of techniques have been invented in order to keep up with newer quantum processors \cite{RMR+06, HS17,SSA16,PGN+17,BIS+17,CZX+18,LWY+18,aaronson2017complexity}.  Unfortunately, this race is one that us classical beings cannot win in the long-term, but there are many good reasons to try. 

Practically speaking, as we broach the boundary of the unsimulable, it is important to verify that our quantum computers are behaving as predicted.  Classical simulations in the regime of NISQ \cite{P18} computers may prove invaluable as an experimental testbed for characterizations of noise, for the development of quantum error correction, and for the verification of quantum systems. Furthermore, classifying the boundary beyond which our quantum computers are doing something genuinely \emph{unattainable} classically is of fundamental interest.  This landmark, termed \emph{quantum supremacy} \cite{BIS+16}, represents the point beyond which we must trust the theory of quantum mechanics rather than our limited simulation of it. It is then important to both our capacity for testing our quantum processors, as well as our pride as classical beings, to push this boundary as far as possible.  Only then will quantum supremacy be meaningful.

Towards this goal, increasing attention has been devoted to general quantum simulation.  Already, there are more than $100$ classical simulators for various types of quantum systems available \footnote{A list of QC simulators can be found at \href{https://quantiki.org/wiki/list-qc-simulators}{https://quantiki.org/wiki/list-qc-simulators.}} . Several different types of simulators \cite{RMR+06, HS17,SSA16,PGN+17,BIS+17,CZX+18,LWY+18} have produced a wide range of results in different contexts.  More specifically, for a quantum circuit with $N$ qubits and depth $d$, there are two major amplitude-wise simulation approaches. The first approach stores the entire state vector in memory, while the second calculates the amplitude $\alpha_x$ for any $N$-bit string $x\in \{0,1\}^N$.

It is not surprising that for the first approach, memory is a major issue. In this context, the largest quantum system that could be classically simulated one decade ago was a 42-qubit one on the J\"{u}lich supercomputer by the Massively Parallel Quantum Computer Simulator \cite{RMR+06}.  In \cite{SSA16}, Intel's qHiPSTER was used more specifically to simulate quantum supremacy circuits of up to $42$ qubits \cite{BIS+16}. In 2017, quantum supremacy circuits of $45$ qubits were simulated on the Cori II supercomputing system using 0.5 petabytes of memory and 8,192 nodes \cite{HS17}. Finally, in 2018, quantum supremacy circuits operating on $7\times 7$ grids of depth $39$ were simulated on the Sunway TaihuLight supercomputer \cite{LWY+18}.

In this work, we focus on computing a single amplitude, which may be far less restrictive on our memory requirements. Towards this end, we consider an algorithm introduced by Markov and Shi \cite{MS08} for tensor network contraction as a template for simulating quantum circuits. This technique is among the very few that are capable of simulating circuits operating on more than 50 qubits to a reasonable depth. By combining this technique with the Feynman path integral, \cite{BIS+17} introduced the undirected graphical model to evaluate circuits with diagonal operators more efficiently. Their work succeeded in simulating $5\times 9$ qubits to depth $40$ and $7\times 8$ circuits to depth $30$ on a single workstation.

\begin{figure}[htb!]
\centering
\includegraphics[width = 0.65\textwidth]{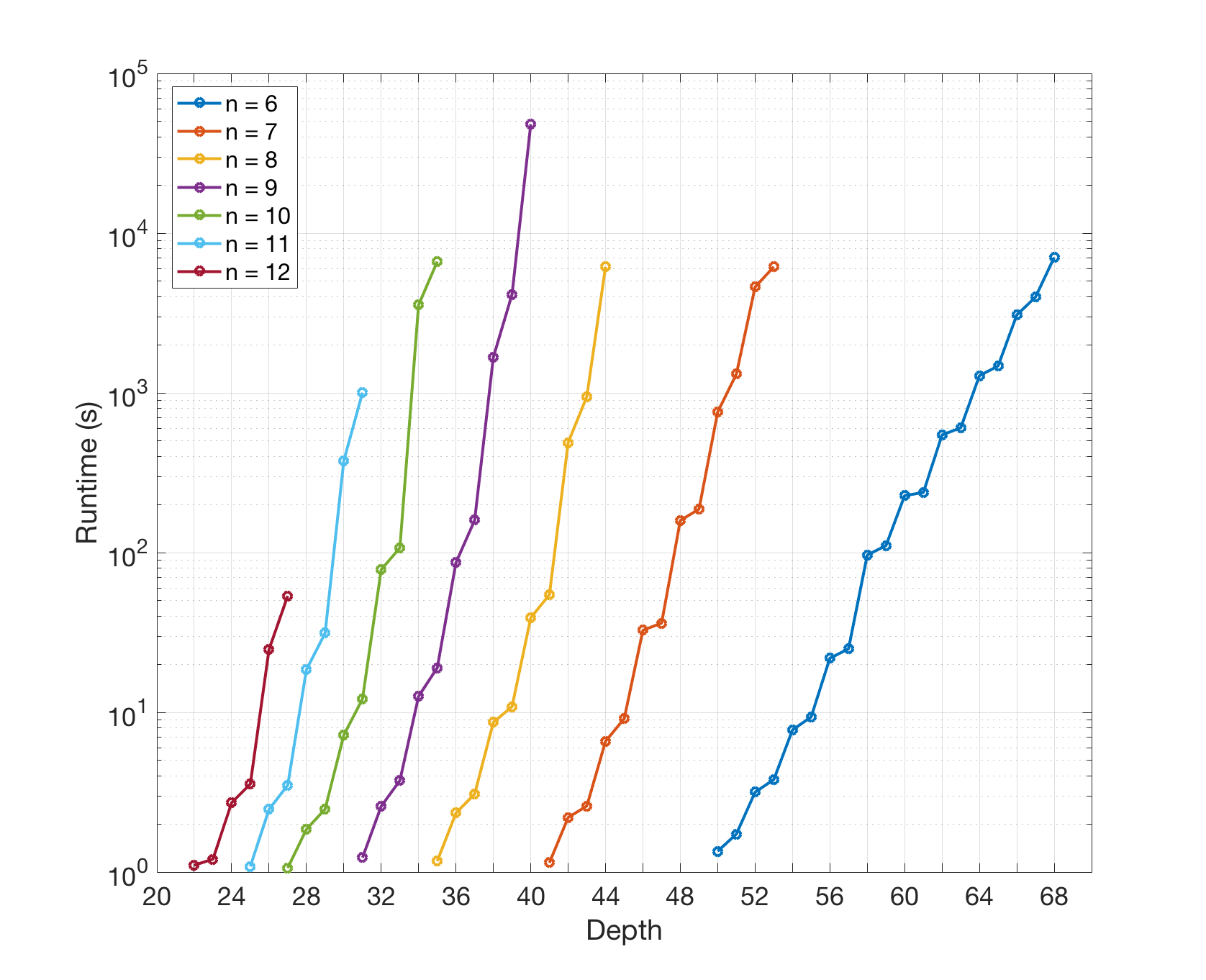}
\caption{The depth vs. runtime plot for our simulator acting on an $n \times n$ square of qubits.  The different lines represent different $n$, and the $y$-axis is plotted on a log-scale.  For reference, two hours of run-time is approximately $10^{3.86}$ seconds.  Although we request 131072 nodes for some instances, when collecting the data, we run one subtask to calculate the running time. This is because the subtasks in the same circuit have the same runtime, and are performed in parallel. Each point represents the $80\%$ percentile runtime among $1000$ randomly generated circuits. We emphasize that the largest depth for $10\times 10$ to $12\times 12$ grids are not the limit of our algorithm. We use QuickBB, which implements an anytime tensor network contraction algorithm, to generate the elimination ordering.  We pre-specify a running time of $60$ seconds, which may not be sufficient for larger depth circuits.  This can be overcome by allowing QuickBB more time to perform its estimation.}
\label{simulation_times}
\end{figure}

Ultimately, this is a numbers game, and so we coarsely summarize existing simulation parameters in Table~\ref{simulators}.  A more detailed presentation of our simulation parameters is shown in Figure~\ref{simulation_times}.

\begin{table}
\centering
 \begin{tabular}{|c||c|c|c|c|} 
 \hline
 Reference & General Technique & Qubits & Depth & \# of Amplitudes \\ 
 \hhline{|=#=|=|=|=|}
 Intel \cite{SSA16} & Full amplitude-vector update & $42$ & High & All \\
 \hline
 ETH \cite{HS17} & Optimized full amplitude-vector update & $5 \times 9$ & 25 & All \\ 
 \hline
 \multirow{2}{*}{IBM \cite{PGN+17}}& \multirow{2}{*}{Tensor-slicing with minimized communication} & $7\times7$ & 27 & All \\

 & & $7 \times 8$ & $23$ & $2^{37}$ out of $2^{56}$ \\
 \hline
 Google \cite{BIS+17} & Preprocessing using undirected graphical model& $7 \times 8$ & 30 & 1 \\
 \hline
 USTC \cite{CZX+18}& Qubit partition with partial vector update & $ 8 \times 9$ & $22$ & 1 \\
 \hline
 \multirow{2}{*}{Sunway \cite{LWY+18}}& \multirow{2}{*}{Dynamic programming qubit partition} & $7 \times 7$ & $39$ & All \\

 & & $7 \times 7$ & $55$ & 1 \\
 \hhline{|=#=|=|=|=|}
 Alibaba & Undirected graphical model with parallelization & $9 \times 9$ & 40 & 1 \\
 \hline
\end{tabular}
 \caption{A very broad overview of existing simulators.  The final column reports the number of amplitudes that are computed by that simulator.}
\label{simulators}
\end{table}
\section{Our Contributions}

In this work, we introduce new techniques for quantum simulation that help us to push this barrier even further.  We implement a single-amplitude simulator that computes $\langle x |\mathcal{C}| 0\cdots 0 \rangle$ for an arbitrary quantum circuit $\mathcal{C}$ and we test our simulator for $\mathcal{C}$ drawn randomly from a restricted circuit class yielding a plausibly intractable sampling problem \cite{BIS+16}.

Our simulation technique is based off of Google's model for variable elimination in the line graph.  This approach is fundamentally tensor network contraction with built-in methods for preprocessing the tensors to minimize cost.  Their work was performed on a single workstation, which is useful for performing individual simulations at low-cost.

In this work, we develop heuristics within this framework that lend themselves naturally to parallelization on a cluster.  We find that, with only a fraction of the power of a cluster, we can perform simulations that break new barriers in the classical simulation of quantum circuits.

Our first contribution is adapting the variable elimination algorithm to the requirements of a typical cluster.  Storing the full amplitude of $50$ qubits requires 16 petabytes of memory if one uses double-precision values. Furthermore, a major restriction is the required inter-node communication. Several attempts have been made to reduce this communication cost, such as global gate optimization \cite{HS17} and qubit reordering \cite{HS17, LWY+18}. In particular, the Alibaba cluster has 10,000 computers, but inter-node communication is very expensive. Thus, the first approach is not suitable for our cluster.

The algorithm we developed is based on tensor network contraction \cite{BIS+17, MS08}, in which treewidth will be the dominant factor in determining the time- and space-complexity. Both of these grow exponentially with the treewidth. The Alibaba cluster has $96 \times 10,000$ CPUs in total, which is just $10\%$ of the CPUs available in the top supercomputer Sunway Taihulight \cite{li2018quantum}. However, the $256$ Sunway processors share $256$ gigabytes of memory, which limits each processor to a smaller amount of memory.  While there are algorithms that limit memory requirements and could take advantage of the powerful Sunway CPUs, our algorithm fits the Alibaba cluster by providing a reasonable number of processors and a reasonable amount of memory for each processor.  

Our second contribution is, naturally, to develop techniques that take full advantage of this parallelization.  We can drastically simplify the graph to be evaluated by parallelizing over the values of key nodes.  This approach balances sequential and parallel techniques to optimize the evaluation of a quantum circuit. The particular ideas that help to achieve our results are the following.

\begin{enumerate}
\item Construct a simplified undirected graph.
\item Divide the whole task into many subtasks while keeping the running time of each subtask as short as possible. We know the running time of each subtask will be exponential in its treewidth. However, calculating treewidth is in NP-hard. The central idea is to use QuickBB's algorithm to roughly estimate the treewidth, and then use the treewidth to estimate the running time.

\emph{Our central observation is that we can remove the most costly nodes by parallelizing over their values.  In spite of its simplicity, we observe tremendous gains using this technique for simulation on a cluster.}
\item Use this estimation again to determine an efficient order of elimination for the remaining nodes in each subtask.
\end{enumerate}

We present a more detailed accounting of these techniques in Section~\ref{thealgorithm}, and the performance of our simulator is summarized in Figure~\ref{simulation_times}. 

The final contribution of this work is to apply this powerful simulator in the context of quantum supremacy. In Section~\ref{nosupremacy}, we give evidence that quantum supremacy is \emph{unachievable} in the low-depth random circuit model defined in \cite{BIS+16}.  We find that this holds true even under idealized (but plausible) target gate fidelities.  This re-emphasizes the integral role error-correction may play in existing quantum computation technologies.

Our paper is outlined as follows.  In Section~\ref{thealgorithm} we detail our algorithm through a simple example, and then describe it in full.  Then, in Section~\ref{numericalresults}, we present our testing parameters as well as the hardware and software used by the algorithm.  In Section~\ref{nosupremacy} we address limitations on showing quantum supremacy through random circuit sampling.  Finally, in Section~\ref{discussion}, we discuss some of the subtleties of our implementation, and potential future considerations.

\section{The Algorithm}\label{thealgorithm}
Our simulator is based on the complex undirected graphical model introduced in \cite{BIS+17}. This is essentially a variant of the tensor network contraction approach proposed by Markov and Shi \cite{MS08}, but which also takes advantage of diagonal gates. We briefly review the undirected graphical model in Subsection~\ref{subsec:undirected}.  We then describe the full algorithm in Subsection~\ref{subsec:algorithm} and explain why our parallelization scheme designed to reduce the treewidth, as this will be the dominant factor in determining the time- and space-complexity.

\subsection{Undirected Graphical Model}\label{subsec:undirected}

In this subsection, we will review the undirected graphical model \cite{BIS+17} which is also used in our base algorithm. 
For any quantum circuit $\mathcal{C}$, the amplitude of a particular bit-string $x$, $\bra{x}\mathcal{C}\ket{0\ldots 0}$, is given by 
\begin{eqnarray*}
\bra{x}\mathcal{C}\ket{0\ldots 0}=\bra{x}\mathcal{C}_d \ldots \mathcal{C}_2 \mathcal{C}_1 \ket{0\ldots 0}
\end{eqnarray*}
where $\mathcal{C}_1$, $\mathcal{C}_2$, $\ldots$, $\mathcal{C}_d$ are unitary matrices corresponding to clock cycles $1, 2, \ldots,$ and $d$, respectively. One may further expand the formula as 
\begin{eqnarray}\label{myequation}
\bra{x}\mathcal{C}\ket{0\ldots 0}=\bra{x}\mathcal{C}_d\ldots \mathcal{C}_2 \mathcal{C}_1 \ket{0\ldots 0}=\sum\limits_{i_1,i_2,\ldots, i_{d-1} \in \{0,1\}^N}\bra{x}\mathcal{C}_d \ket{i_{d-1}} \ldots \bra{i_2} \mathcal{C}_2 \ket{i_1}\bra{i_1}\mathcal{C}_1 \ket{0\ldots 0} .
\label{feynmanpathintegral}
\end{eqnarray}

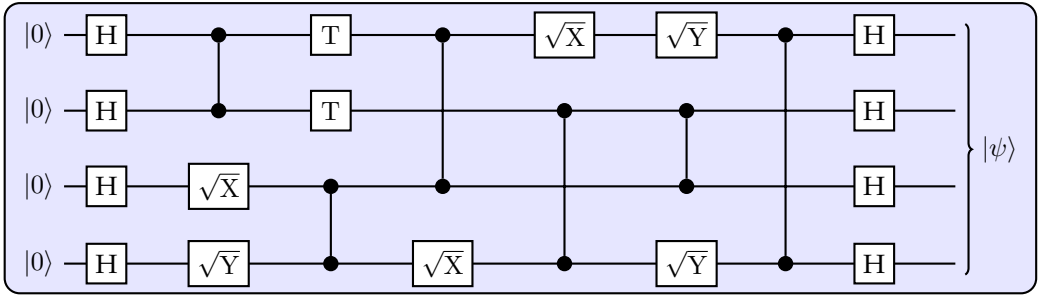
\begin{figure}
    \begin{tikzpicture}[thick]
    % `operator' will only be used by Hadamard (H) gates here.
    % `phase' is used for controlled phase gates (dots).
    % `surround' is used for the background box.
    \tikzstyle{operator} = [draw,fill=white,minimum size=1.5em] 
    \tikzstyle{phase} = [draw,fill,shape=circle,minimum size=5pt,inner sep=0pt]
    \tikzstyle{surround} = [fill=blue!10,thick,draw=black,rounded corners=2mm]
    \matrix[row sep=0.4cm, column sep=0.8cm] (circuit) {
    % First row.
    \node (q1) {\ket{0}}; &[-0.5cm] 
    \node[operator] (H11) {H}; &
    \node[phase] (P12) {}; &
    \node[operator] (T13) {$\mathrm{T}$}; &
    \node[phase] (P14) {}; &
    \node[operator] (X15) {$\mathrm{\sqrt{X}}$}; &
    \node[operator] (Y16){$\mathrm{\sqrt{Y}}$}; &
    \node[phase] (P17) {}; &
    \node[operator] (H18) {H}; &
    \coordinate (end1); \\
    % Second row.
    \node (q2) {\ket{0}}; &
    \node[operator] (H21) {H}; &
    \node[phase] (P22) {}; &
    \node[operator] (T23) {$\mathrm{T}$}; &    
    &
    \node[phase] (P25) {}; &
    \node[phase] (P26){}; &
    &
    \node[operator] (H28) {H}; &
    \coordinate (end2);\\
    % Third row.
    \node (q3) {\ket{0}}; &
    \node[operator] (H31) {H}; &
    \node[operator] (X32) {$\sqrt{\mathrm{X}}$}; &
    \node[phase] (P33) {}; &
    \node[phase] (P34) {}; &
    &
    \node[phase] (P36){}; &
    &
    \node[operator] (H38) {H}; &
    \coordinate (end3); \\
    % Fouth row.
    \node (q4) {\ket{0}}; &
    \node[operator] (H41) {H}; &
    \node[operator] (T42) {$\sqrt{\mathrm{Y}}$}; &
    \node[phase] (P43) {}; &
    \node[operator] (X44) {$\mathrm{\sqrt{X}}$}; &
    \node[phase] (P45) {}; & 
    \node[operator] (P46){$\mathrm{\sqrt{Y}}$}; &
    \node[phase] (P47) {}; &
    \node[operator] (H48) {H}; &
    \coordinate (end4); \\
    };
    % Draw bracket on right with resultant state.
    \draw[decorate,decoration={brace},thick]
        ($(circuit.north east)-(0cm,0.3cm)$)
        to node[midway,right] (bracket) {$\textrm{\ }\displaystyle\ket{\psi}$}
        ($(circuit.south east)+(0cm,0.3cm)$);
    \begin{pgfonlayer}{background}
        % Draw background box.
        \node[surround] (background) [fit = (q1) (H48) (bracket)] {};
        % Draw lines.
        \draw[thick] (q1) -- (end1)  (q2) -- (end2) (q3) -- (end3) (q4) -- (end4) (P12) -- (P22) (P33) -- (P43) (P14) -- (P34) (P26) -- (P36) (P25) -- (P45) (P17) -- (P47);
    \end{pgfonlayer}
    \end{tikzpicture}
\caption{The example circuit $\cal{C}$ that we evaluate using the graphical model.} \label{fig:circuit}
\end{figure}

For example, let's consider a circuit with qubit number $N=4$ and depth $d=8$ as described in Figure~\ref{fig:circuit}. Consider $\mathcal{C} = \mathcal{C}_8 \ldots \mathcal{C}_2\mathcal{C}_1$ defined by

\begin{minipage}{.45\textwidth}
\begin{align*}
\mathcal{C}_1 &= \mathrm{H}_1\otimes \mathrm{H}_2\otimes \mathrm{H}_3 \otimes \mathrm{H}_4;\\ 
\mathcal{C}_2 &= \mathrm{CZ}_{1,2}\otimes \sqrt{\mathrm{X}}_{3}\otimes \sqrt{\mathrm{Y}}_{4}; \\ 
\mathcal{C}_3 &= \mathrm{T}_1\otimes \mathrm{T}_2 \otimes \mathrm{CZ}_{3,4}; \\ 
\mathcal{C}_4 &= \mathrm{CZ}_{1,3}\otimes \mathrm{I}_2 \otimes \sqrt{\mathrm{X}}_4; 
\end{align*}
\end{minipage}
\begin{minipage}{.45\textwidth}
\begin{align*}
\mathcal{C}_5 &= \mathrm{CZ}_{2,4}\otimes \mathrm{I}_3 \otimes \sqrt{\mathrm{X}}_1; \\ 
\mathcal{C}_6 &= \mathrm{CZ}_{2,3}\otimes \sqrt{\mathrm{Y}}_1 \otimes \sqrt{\mathrm{Y}}_4; \\ 
\mathcal{C}_7 &= \mathrm{CZ}_{1,4}\otimes \mathrm{I}_{2,3} ;\\ 
\mathcal{C}_8 &= \mathrm{H}_1\otimes \mathrm{H}_2\otimes \mathrm{H}_3 \otimes \mathrm{H}_4.
\end{align*}
\end{minipage}

\medskip
Then we may expand Equation~\ref{myequation} as,
\begin{eqnarray}\label{eq:feynman}
\sum\limits_{i_1,i_2, i_3,i_4, i_5, i_6,i_7 \in \{0,1\}^4}  && \bra{x}\mathrm{H}_1\otimes \mathrm{H}_2\otimes \mathrm{H}_3 \otimes \mathrm{H}_4\ket{i_7}\bra{i_7}\mathrm{CZ}_{1,4}\otimes \mathrm{I}_{2,3} \ket{i_6} \nonumber\\
&\cdot& \bra{i_6}   \mathrm{CZ}_{2,3}\otimes \sqrt{\mathrm{Y}}_1 \otimes \sqrt{\mathrm{Y}}_4 \ket{i_5} \bra{i_5} \mathrm{CZ}_{2,4}\otimes \mathrm{I}_3 \otimes \sqrt{\mathrm{X}}_1 \ket{i_4}\nonumber\\
&\cdot&\bra{i_4} \mathrm{CZ}_{1,3}\otimes \mathrm{I}_2 \otimes \sqrt{\mathrm{X}}_4 \ket{i_3} 
\bra{i_3} \mathrm{T}_1\otimes \mathrm{T}_2 \otimes \mathrm{CZ}_{3,4}\ket{i_2} \nonumber \\
&\cdot&  \bra{i_2}  \mathrm{CZ}_{1,2}\otimes \sqrt{\mathrm{X}}_{3}\otimes \sqrt{\mathrm{Y}}_{4}\ket{i_1}
\bra{i_1}\mathrm{H}_1\otimes \mathrm{H}_2\otimes \mathrm{H}_3 \otimes \mathrm{H}_4 \ket{0000}.  
\end{eqnarray}

\vskip 0.5 cm

The summation in Equation~\ref{eq:feynman} is carried out over the seven $4$-bit strings $i_1,i_2, \ldots, i_7$. Because $\mathrm{T}$ and $\mathrm{CZ}$ are diagonal matrices, the terms in summation will appear only if $i_1^{(1,2)}=i_2^{(1,2)}$, $i_2=i_3$, $i_3^{(123)}=i_4^{(123)}$, $i_4^{(234)}=i_5^{(234)}$, $i_5^{(2,3)}=i_6^{(2,3)}$ and $i_6=i_7$. By considering the diagonal gates in this way, we have drastically reduced the total number of terms from $2^{28}$ to $2^{10}$.

We now formulate the above procedure in the language of undirected graphical models. Given the index sequences $(i_0=0\cdots 0, i_1,i_2, \ldots, i_{d-1}, i_d=x)$ in Equation~\ref{feynmanpathintegral}, we construct a graph $G$, where each distinct variable $i_k^{(j)}$ corresponds to a vertex, and two vertices are connected by an edge if there is an operator acting on both of them. Each term in the Feynman path integral then corresponds to a complex number associated to labelling all vertices in the graph by $\{0,1\}$. 

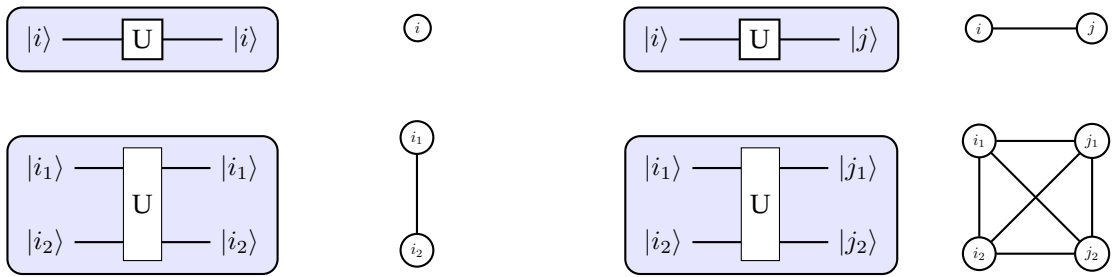
\begin{figure}[h]
\begin{minipage}{.25\textwidth}
\begin{center}
    \begin{tikzpicture}[thick]
    % `operator' will only be used by Hadamard (H) gates here.
    % `phase' is used for controlled phase gates (dots).
    % `surround' is used for the background box.
    \tikzstyle{operator} = [draw,fill=white,minimum size=1.5em] 
    \tikzstyle{phase} = [draw,fill,shape=circle,minimum size=5pt,inner sep=0pt]
    \tikzstyle{surround} = [fill=blue!10,thick,draw=black,rounded corners=2mm]
    \matrix[row sep=0.4cm, column sep=0.8cm] (circuit) {
    % First row.
    \node (input) {\ket{i}}; &
    \node[operator] (U) {U}; & \node (output) {$\ket{i}$};\\
    };
    % Draw bracket on right with resultant state.
    \begin{pgfonlayer}{background}
        % Draw background box.
        \node[surround] (background) [fit = (input) (U) (output)] {};
        % Draw lines.
        \draw[thick] (input) -- (output);
    \end{pgfonlayer}
    \end{tikzpicture}

   \vskip 0.8cm

    \begin{tikzpicture}[thick]
    % `operator' will only be used by Hadamard (H) gates here.
    % `phase' is used for controlled phase gates (dots).
    % `surround' is used for the background box.
    \tikzstyle{operator} = [draw,fill=white,minimum size=1.5em] 
    \tikzstyle{operator2}= [draw,fill=white,minimum height=1.5cm]
    \tikzstyle{phase} = [draw,fill,shape=circle,minimum size=5pt,inner sep=0pt]
    \tikzstyle{surround} = [fill=blue!10,thick,draw=black,rounded corners=2mm]
    \matrix[row sep=0.4cm, column sep=0.8cm] (circuit) {
    % First row.
    \node (input1) {\ket{i_1}}; &
    \node[] (U21) {}; & \node (output1) {$\ket{i_1}$};\\
    % Second row.
    \node (input2) {\ket{i_2}}; &
    \node[] (U22) {}; & \node (output2) {$\ket{i_2}$};\\
    };
    % Draw bracket on right with resultant state.

    % Draw bracket on right with resultant state.
    \begin{pgfonlayer}{background}
        % Draw background box.
        \node[surround] (background) [fit = (input1) (U22) (output2)] {};
        % Draw lines.
        \draw[thick] (input1) -- (output1)  (input2) -- (output2);
    \node[operator2] at ($(U21.south)+(0cm,-0.35cm)$){U};  %<-- for large U
    \end{pgfonlayer}
    \end{tikzpicture}

\end{center}
\end{minipage}
\hspace{-1cm}
\begin{minipage}{.25\textwidth}
\begin{center}

\begin{tikzpicture}[auto, scale=0.6, node distance=2.5cm, every loop/.style={scale=0.6}, 
                    thick,main node/.style={scale=0.6, circle,draw,font=\sffamily\bfseries}]
    \tikzstyle{surround} = [fill=blue!10,thick,draw=black,rounded corners=2mm]

  \node[main node] (1) [above right]{$i$};
\end{tikzpicture}

   \vskip 1cm

\begin{tikzpicture}[auto, scale=0.6, node distance=2.5cm, every loop/.style={scale=0.6}, 
                    thick,main node/.style={scale=0.6, circle,draw,font=\sffamily\bfseries}]
    \tikzstyle{surround} = [fill=blue!10,thick,draw=black,rounded corners=2mm]

  \node[main node] (1) {$i_1$};
  \node[main node] (2) [below  of=1] {$i_2$};
  \path[every node/.style={font=\sffamily\small}]
    (1) edge node [yshift=0pt] {} (2);
\end{tikzpicture}

\end{center}
\end{minipage}
\begin{minipage}{.25\textwidth}
\begin{center}

    \begin{tikzpicture}[thick]
    % `operator' will only be used by Hadamard (H) gates here.
    % `phase' is used for controlled phase gates (dots).
    % `surround' is used for the background box.
    \tikzstyle{operator} = [draw,fill=white,minimum size=1.5em] 
    \tikzstyle{phase} = [draw,fill,shape=circle,minimum size=5pt,inner sep=0pt]
    \tikzstyle{surround} = [fill=blue!10,thick,draw=black,rounded corners=2mm]
    \matrix[row sep=0.4cm, column sep=0.8cm] (circuit) {
    % First row.
    \node (input) {\ket{i}}; &
    \node[operator] (U) {U}; & \node (output) {$\ket{j}$};\\
    };
    % Draw bracket on right with resultant state.
    \begin{pgfonlayer}{background}
        % Draw background box.
        \node[surround] (background) [fit = (input) (U) (output)] {};
        % Draw lines.
        \draw[thick] (input) -- (output);
    \end{pgfonlayer}
    \end{tikzpicture}
    
    \vskip 0.8cm

    \begin{tikzpicture}[thick]
    % `operator' will only be used by Hadamard (H) gates here.
    % `phase' is used for controlled phase gates (dots).
    % `surround' is used for the background box.
    \tikzstyle{operator} = [draw,fill=white,minimum size=1.5em] 
    \tikzstyle{operator2}= [draw,fill=white,minimum height=1.5cm]
    \tikzstyle{phase} = [draw,fill,shape=circle,minimum size=5pt,inner sep=0pt]
    \tikzstyle{surround} = [fill=blue!10,thick,draw=black,rounded corners=2mm]
    \matrix[row sep=0.4cm, column sep=0.8cm] (circuit) {
    % First row.
    \node (input1) {\ket{i_1}}; &
    \node[] (U21) {}; & \node (output1) {$\ket{j_1}$};\\
    % Second row.
    \node (input2) {\ket{i_2}}; &
    \node[] (U22) {}; & \node (output2) {$\ket{j_2}$};\\
    };
    % Draw bracket on right with resultant state.

    % Draw bracket on right with resultant state.
    \begin{pgfonlayer}{background}
        % Draw background box.
        \node[surround] (background) [fit = (input1) (U22) (output2)] {};
        % Draw lines.
        \draw[thick] (input1) -- (output1)  (input2) -- (output2);
    \node[operator2] at ($(U21.south)+(0cm,-0.35cm)$){U};  %<-- for large U
    \end{pgfonlayer}
    \end{tikzpicture}
\end{center}
\end{minipage}
\hspace{-1cm}
\begin{minipage}{.25\textwidth}
\begin{center}

\begin{tikzpicture}[auto, scale=0.6, node distance=2.5cm, every loop/.style={scale=0.6}, 
                    thick,main node/.style={scale=0.6, circle,draw,font=\sffamily\bfseries}]
    \tikzstyle{surround} = [fill=blue!10,thick,draw=black,rounded corners=2mm]

  \node[main node] (1) {$i$};
  \node[main node] (2) [right  of=1] {$j$};
  \path[every node/.style={font=\sffamily\small}]
    (1) edge node [yshift=0pt] {} (2);
\end{tikzpicture}

   \vskip 1cm

\begin{tikzpicture}[auto, scale=0.6, node distance=2.5cm, every loop/.style={scale=0.6}, 
                    thick,main node/.style={scale=0.6, circle,draw,font=\sffamily\bfseries}]
    \tikzstyle{surround} = [fill=blue!10,thick,draw=black,rounded corners=2mm]

  \node[main node] (1) {$i_1$};
  \node[main node] (2) [below  of=1] {$i_2$};
  \node[main node] (3) [right  of=1] {$j_1$};
  \node[main node] (4) [below  of=3] {$j_2$};
  \path[every node/.style={font=\sffamily\small}]
    (1) edge node [yshift=0pt] {} (2)
    (1) edge node [yshift=0pt] {} (3)
    (1) edge node [yshift=0pt] {} (4)
    (2) edge node [yshift=0pt] {} (3)
    (2) edge node [yshift=0pt] {} (4)
    (3) edge node [yshift=0pt] {} (4);
\end{tikzpicture}
\end{center}
\end{minipage}

\caption{Single- and two-qubit diagonal and non-diagonal gates and their corresponding graphical gadgets.}
\label{tensornodes}
\end{figure}

Such a graph can be simplified if some tensor operators happen to be diagonal. For example, if node $u$ and $v$ are connected by a sigle-qubit diagonal gate, then one term in the Feynman path integral can only survive if the corresponding labelling we choose satisfies $u=v$. Therefore, the two nodes $u$ and $v$ can be merged together. Similarly, the tensors on the lefthand side of Figure~\ref{tensornodes} correspond to the graph gadgets on the right.

\begin{figure}
\begin{tikzpicture}[auto, scale=0.6, node distance=2.5cm, every loop/.style={scale=0.6}, 
                    thick,main node/.style={scale=0.6, circle,draw,font=\sffamily\bfseries}, minimum size = 1cm]
    \tikzstyle{surround} = [fill=blue!10,thick,draw=black,rounded corners=2mm]
  \node[main node, draw=none] (v1) []{};    
  \node[main node] (5) [right of = v1] {$a$};
  \node[main node, draw=none] (v2) [below of =v1]{};    
  \node[main node, draw=none] (v3) [right of =v2]{};    
  \node[main node] (6) [right of=v3] {$b$};
  \node[main node] (7) [below of=v2] {$c$};
  \node[main node] (8) [below of=7] {$d$};
  \node[main node] (9) [right of=7] {$e$};
  \node[main node] (10) [right of=5] {$h$};  
  \node[main node] (11) [right of=10] {$i$};  
  \node[main node] (12) [right of=8] {$f$};
  \node[main node] (13) [right of=12] {$g$};
  \node[main node] (14) [right of=13] {$j$};
  \path[every node/.style={font=\sffamily\small}]
    (6) edge node [] {} (9)     
    (6) edge node [] {} (13)     
    (12) edge node [] {} (8)     
    (9) edge node [] {} (12)     
    (7) edge node [] {} (9)     
    (12) edge node [] {} (13)         
    (5) edge node [] {} (9)         
    (5) edge node [] {} (10)         
    (11) edge node [] {} (10)         
    (11) edge node [] {} (14)         
    (13) edge node [] {} (14)         
    (5) edge node [] {} (6)     ;
    \begin{pgfonlayer}{background}
        % Draw background box.
        \node[surround] (background) [fit = (v1) (14) ] {};
        % Draw lines.
    \end{pgfonlayer}       
\end{tikzpicture}
\caption{The undirected graphical model associated with the circuit in Figure~\ref{fig:circuit}.} \label{fig:graph}
\end{figure}
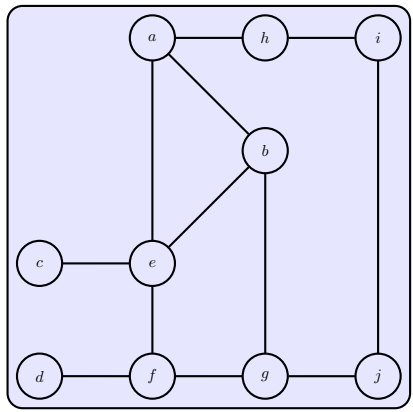

Back to our example in Figure~\ref{fig:circuit}, let $i_1=abcd$, $i_2=i_3=abef$, $i_4=abeg$, $i_5=hbeg$, $i_6=i_7=ibej$. Then the corresponding undirected graph can be drawn as in Figure~\ref{fig:graph}.

\subsection{Our Algorithm}\label{subsec:algorithm}

\subsubsection{Variable elimination}\label{subsec:elimination}
Our base algorithm for evaluating the sum in Equation~\ref{myequation} is similar to the algorithm in \cite{BIS+17}, and proceeds by eliminating one variable at a time. This process will usually produce tensors of rank greater than 2. To eliminate the binary variable $v = i_k^{(j)}$ from Equation~\ref{myequation}:
\begin{enumerate}
\item Find the set of tensors $T_v = \{\tau \mid v\text{ appears in }\tau\}$, and the set of variables $V_v = \bigcup_{\tau \in T_v} \{v' \mid v'\text{ appears in }\tau\}$.
\item Multiply together all tensors $\tau \in T_v$ to obtain a new tensor $\sigma$ of rank $|V_v|$ and indexed by variables in $|V_v|$.
\item Sum $\sigma$ over the index corresponding to $v$ to obtain $\sigma'$.
\item Remove the variable $v$ and all the tensors in $T_v$ from the summation, and then add the new tensor $\sigma'$. Update the undirected graph by connecting any two neighbors of vertex $v$, and then removing $v$.
\end{enumerate}

To evaluate the entire sum, repeat the above steps to eliminate all of the variables in any convenient order. When all variables are eliminated, we get a tensor of rank $0$ (i.e.\ a complex number) which is exactly the value of the amplitude.

Steps 2 and 3 constitute the bulk of the computation. Fortunately, they can be computed on Python using the \texttt{numpy} package with a single call to the function \texttt{numpy.einsum}. This is optimized by first combining the small input tensors, and then only evaluating large tensors in the final step \footnote{This optimization was first implemented as a third party add-on package,  \texttt{opt\_einsum}. Afterwards, it was added to \texttt{numpy} in version 1.12.0, and became turned on by default since \texttt{numpy} v1.14.0.}. 

If eliminating a variable $v$ incurs a noticeable time cost, then usually the dominant term in that cost comes from the appearance of a high rank tensor $\sigma'$. Therefore, we estimate that the time cost of each step by the size of $\sigma'$. Given the undirected graph at that step, the size of $\sigma'$ is $2^{\deg(v)}$. The cost function of a particular contraction ordering is then the sum of the costs over all steps. We will use this estimation as a guide when simplifying the graph by fixing the values of certain variables, as we describe next.

\subsubsection{Parallelization by fixing the values of variables} \label{parallel}

There is an even more straightforward method to evaluate Equation~\ref{myequation}, which is just to split the sum into pieces. We can simply choose any variable $v$ and evaluate the summation twice, once with the value of $v$ fixed to 0 and once with the value of $v$ fixed to 1, and then combine the outcomes.

Similar to eliminating a variable, fixing the value of a variable \emph{also} removes it from the summation. However, whereas eliminating a variable usually amalgamates several tensors into a higher rank tensor, fixing the value of a variable does not introduce any new tensor. This can potentially simplify the summation, as illustrated in Figure~\ref{fig:para}. In the undirected graph model, fixing the value of a variable translates to removing the corresponding vertex along with all of its edges.

\begin{center}
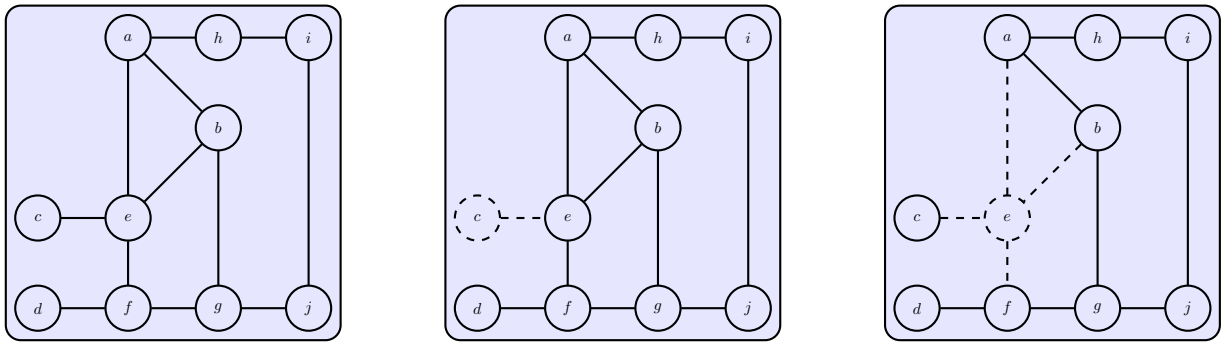
\begin{figure}[h]
\begin{tikzpicture}[auto, scale=0.45, node distance = 2cm,
                    thick,main node/.style={scale = 0.6, circle,draw,font=\sffamily\bfseries}, minimum size = 1cm]
    \tikzstyle{surround} = [fill=blue!10,thick,draw=black,rounded corners=2mm]
      \node[main node, draw=none] (v1) []{};    
  \node[main node] (5) [right of = v1] {$a$};
  \node[main node, draw=none] (v2) [below of =v1]{};    
  \node[main node, draw=none] (v3) [right of =v2]{};    
  \node[main node] (6) [right of=v3] {$b$};
  \node[main node] (7) [below of=v2] {$c$};
  \node[main node] (8) [below of=7] {$d$};
  \node[main node] (9) [right of=7] {$e$};
  \node[main node] (10) [right of=5] {$h$};  
  \node[main node] (11) [right of=10] {$i$};  
  \node[main node] (12) [right of=8] {$f$};
  \node[main node] (13) [right of=12] {$g$};
  \node[main node] (14) [right of=13] {$j$};
  \path[every node/.style={font=\sffamily\small}]
    (6) edge node [] {} (9)     
    (6) edge node [] {} (13)     
    (12) edge node [] {} (8)     
    (9) edge node [] {} (12)     
    (7) edge node [] {} (9)     
    (12) edge node [] {} (13)         
    (5) edge node [] {} (9)         
    (5) edge node [] {} (10)         
    (11) edge node [] {} (10)         
    (11) edge node [] {} (14)         
    (13) edge node [] {} (14)         
    (5) edge node [] {} (6)     ;
    \begin{pgfonlayer}{background}
        % Draw background box.
        \node[surround] (background) [fit = (v1) (14)] {};
        % Draw lines.
    \end{pgfonlayer}       
\end{tikzpicture}
\qquad % <----------------- SPACE BETWEEN PICTURES
\begin{tikzpicture}[auto, scale=0.45, node distance = 2cm,
                    thick,main node/.style={scale = 0.6, circle,draw,font=\sffamily\bfseries}, minimum size = 1cm]
    \tikzstyle{surround} = [fill=blue!10,thick,draw=black,rounded corners=2mm]
  \node[main node, draw=none] (v1) []{};    
  \node[main node] (5) [right of = v1] {$a$};
  \node[main node, draw=none] (v2) [below of =v1]{};    
  \node[main node, draw=none] (v3) [right of =v2]{};    
  \node[main node] (6) [right of=v3] {$b$};
  \node[main node,dashed] (7) [below of=v2] {$c$};
  \node[main node] (8) [below of=7] {$d$};
  \node[main node] (9) [right of=7] {$e$};
  \node[main node] (10) [right of=5] {$h$};  
  \node[main node] (11) [right of=10] {$i$};  
  \node[main node] (12) [right of=8] {$f$};
  \node[main node] (13) [right of=12] {$g$};
  \node[main node] (14) [right of=13] {$j$};
  \path[every node/.style={font=\sffamily\small}]
    (6) edge node [] {} (9)     
    (6) edge node [] {} (13)     
    (9) edge node [] {} (12)     
    (12) edge node [] {} (13)         
    (5) edge node [] {} (9)         
    (5) edge node [] {} (10)         
    (11) edge node [] {} (10)         
    (11) edge node [] {} (14)         
    (13) edge node [] {} (14)         
    (5) edge node [] {} (6)
    (12) edge node [] {} (8)     ;    
    \path[every node/.style={font=\sffamily\small}, dashed]
    (7) edge node [] {} (9)     ;
    \begin{pgfonlayer}{background}
        % Draw background box.
        \node[surround] (background) [fit = (v1) (14)] {};
        % Draw lines.
    \end{pgfonlayer}       
\end{tikzpicture}
\qquad % <----------------- SPACE BETWEEN PICTURES
\begin{tikzpicture}[auto, scale=0.45, node distance = 2cm,
                    thick,main node/.style={scale = 0.6, circle,draw,font=\sffamily\bfseries}, minimum size = 1cm]
    \tikzstyle{surround} = [fill=blue!10,thick,draw=black,rounded corners=2mm]
  \node[main node, draw=none] (v1) []{};    
  \node[main node] (5) [right of = v1] {$a$};
  \node[main node, draw=none] (v2) [below of =v1]{};    
  \node[main node, draw=none] (v3) [right of =v2]{};    
  \node[main node] (6) [right of=v3] {$b$};
  \node[main node] (7) [below of=v2] {$c$};
  \node[main node] (8) [below of=7] {$d$};
  \node[main node, dashed] (9) [right of=7] {$e$};
  \node[main node] (10) [right of=5] {$h$};  
  \node[main node] (11) [right of=10] {$i$};  
  \node[main node] (12) [right of=8] {$f$};
  \node[main node] (13) [right of=12] {$g$};
  \node[main node] (14) [right of=13] {$j$};
  \path[every node/.style={font=\sffamily\small}]
    (6) edge node [] {} (13)     
    (12) edge node [] {} (8)     
    (12) edge node [] {} (13)         
    (5) edge node [] {} (10)         
    (11) edge node [] {} (10)         
    (11) edge node [] {} (14)         
    (13) edge node [] {} (14)         
    (5) edge node [] {} (6) ;
    \path[every node/.style={font=\sffamily\small}, dashed]
    (5) edge node [] {} (9)         
    (9) edge node [] {} (12)     
    (6) edge node [] {} (9)     
    (7) edge node [] {} (9)    
    ;
    \begin{pgfonlayer}{background}
        % Draw background box.
        \node[surround] (background) [fit = (v1) (14)] {};
        % Draw lines.
    \end{pgfonlayer}       
\end{tikzpicture}

\caption{Illustration of parallelizing by fixing the value of a node. The initial undirected graphical model is shown in the leftmost figure.  Fixing the value of node $c$ does not help simplify the graph much, and is displayed in the middle figure. Instead fixing the value of node $e$ simplifies the graph significantly, and is shown in the rightmost figure.}\label{fig:para}
\end{figure}
\end{center}

On the other hand, evaluating an undirected graph using only this strategy would be highly inefficient compared to our base algorithm.  This is because each time we remove a vertex in this manner, we need to recursively evaluate the resulting graph twice. The number of evaluations blows up exponentially in the number of vertices we remove this way.

The advantage of this strategy is that \emph{all of these evaluations can be done in parallel}. Our overall strategy is to remove $t$ vertices this way, essentially dividing the graph evaluation task into $2^t$ subtasks, and then perform each subtask using the base algorithm. Of course, this is only effective if the cost of evaluating the resulting graph is substantially lower than the cost of evaluating the original graph.

Removing different vertices may have wildly different effects on the cost of the base algorithm. Intuitively, removing high-degree vertices should simplify the graph further, but even a low-degree vertex can be a ``key vertex" that holds large parts of the graph together. % Maybe we should give an example here... Alternatively, "it can be unclear what the choice should be among a large number of vertices with the same degree".
In order to choose which vertices to remove, we use a greedy strategy based on the estimated time cost of the base algorithm.

Finally, since our goal is only to compute the value of $\bra{x}\mathcal{C}\ket{0\ldots 0}$, the final values of all qubits are fixed to $x$, and so all corresponding vertices can be removed for free \footnote{The vertices corresponding to the initial values of qubits could also be removed for free, but for the family of circuits we consider, these vertices are all degree $1$ and so the benefit would be negligible.}. Fortunately, for the circuits we consider, we find that this optimization simplifies the initial graph \emph{considerably}. % Should give data of treewidth and/or estimated time cost.

\subsubsection{The variable elimination ordering}

Similar to \cite{BIS+17}, the variable elimination ordering can heavily affect the time-complexity of our base algorithm (See Fig.~\ref{fig:order} for an example). Additionally, it can also affect the choice of vertices we remove before applying our base algorithm, as this choice is based on the estimated time cost to evaluate the graph.

In \cite{BIS+17},  all of the TensorFlow experiments use the "vertical ordering", eliminating variables corresponding to each qubit in temporal order before moving on to the next qubit. In the same paper, the heuristic algorithm QuickBB \cite{VR04} is also used to find a better elimination ordering (and a better upper bound for the treewidth). We observe that QuickBB usually finds an ordering that improves both the maximum tensor rank \emph{and} the estimated time cost. Furthermore, initially using QuickBB makes the greedy vertex removal more efficient in reducing the time cost. 

However, QuickBB itself is slow enough that we cannot afford to run it whenever we would like to estimate the time cost. To compromise, we run QuickBB once initially for each circuit to obtain an elimination ordering, and then use that ordering throughout to remove vertices. We then run QuickBB a second time after vertex removal to obtain a new elimination ordering that will evaluate the graph efficiently.

% Below may be too much information, please feel free to delete any or all of them

In our experiments, we use an off-the-shelf QuickBB implementation by \cite{VR04}. We specified the command line options \texttt{-{}-min-fill-ordering} to perform branching using min-fill ordering, and \texttt{-{}-time 60} to run QuickBB for 60 seconds both before and after the greedy vertex removal.

\begin{center}
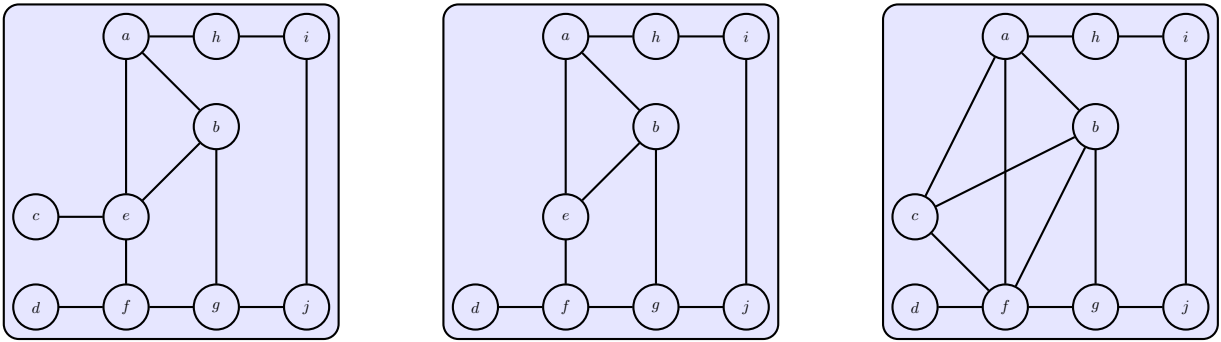
\begin{figure}[h]
\begin{tikzpicture}[auto, scale=0.45, node distance = 2cm,
                    thick,main node/.style={scale = 0.6, circle,draw,font=\sffamily\bfseries}, minimum size = 1cm]
    \tikzstyle{surround} = [fill=blue!10,thick,draw=black,rounded corners=2mm]
  \node[main node, draw=none] (v1) []{};    
  \node[main node] (5) [right of = v1] {$a$};
  \node[main node, draw=none] (v2) [below of =v1]{};    
  \node[main node, draw=none] (v3) [right of =v2]{};    
  \node[main node] (6) [right of=v3] {$b$};
  \node[main node] (7) [below of=v2] {$c$};
  \node[main node] (8) [below of=7] {$d$};
  \node[main node] (9) [right of=7] {$e$};
  \node[main node] (10) [right of=5] {$h$};  
  \node[main node] (11) [right of=10] {$i$};  
  \node[main node] (12) [right of=8] {$f$};
  \node[main node] (13) [right of=12] {$g$};
  \node[main node] (14) [right of=13] {$j$};
  \path[every node/.style={font=\sffamily\small}]
    (6) edge node [] {} (9)     
    (6) edge node [] {} (13)     
    (12) edge node [] {} (8)     
    (9) edge node [] {} (12)     
    (7) edge node [] {} (9)     
    (12) edge node [] {} (13)         
    (5) edge node [] {} (9)         
    (5) edge node [] {} (10)         
    (11) edge node [] {} (10)         
    (11) edge node [] {} (14)         
    (13) edge node [] {} (14)         
    (5) edge node [] {} (6)     ;
    \begin{pgfonlayer}{background}
        % Draw background box.
        \node[surround] (background) [fit = (v1) (14)] {};
        % Draw lines.
    \end{pgfonlayer}       
\end{tikzpicture}
\qquad % <----------------- SPACE BETWEEN PICTURES
\begin{tikzpicture}[auto, scale=0.45, node distance = 2cm,
                    thick,main node/.style={scale = 0.6, circle,draw,font=\sffamily\bfseries}, minimum size = 1cm]
    \tikzstyle{surround} = [fill=blue!10,thick,draw=black,rounded corners=2mm]
  \node[main node, draw=none] (v1) []{};    
  \node[main node] (5) [right of = v1] {$a$};
  \node[main node, draw=none] (v2) [below of =v1]{};    
  \node[main node, draw=none] (v3) [right of =v2]{};    
  \node[main node] (6) [right of=v3] {$b$};
  \node[main node, draw = none] (7) [below of=v2] {};
  \node[main node] (8) [below of=7] {$d$};
  \node[main node] (9) [right of=7] {$e$};
  \node[main node] (10) [right of=5] {$h$};  
  \node[main node] (11) [right of=10] {$i$};  
  \node[main node] (12) [right of=8] {$f$};
  \node[main node] (13) [right of=12] {$g$};
  \node[main node] (14) [right of=13] {$j$};
  \path[every node/.style={font=\sffamily\small}]
    (6) edge node [] {} (9)     
    (6) edge node [] {} (13)     
    (9) edge node [] {} (12)     
    (12) edge node [] {} (13)         
    (5) edge node [] {} (9)         
    (5) edge node [] {} (10)         
    (11) edge node [] {} (10)         
    (11) edge node [] {} (14)         
    (13) edge node [] {} (14)         
    (5) edge node [] {} (6)     
    (12) edge node [] {} (8)     ;    
    \begin{pgfonlayer}{background}
        % Draw background box.
        \node[surround] (background) [fit = (v1) (14)] {};
        % Draw lines.
    \end{pgfonlayer}       
\end{tikzpicture}
\qquad % <----------------- SPACE BETWEEN PICTURES
\begin{tikzpicture}[auto, scale=0.45, node distance = 2cm,
                    thick,main node/.style={scale = 0.6, circle,draw,font=\sffamily\bfseries}, minimum size = 1cm]
    \tikzstyle{surround} = [fill=blue!10,thick,draw=black,rounded corners=2mm]
  \node[main node, draw=none] (v1) []{};    
  \node[main node] (5) [right of = v1] {$a$};
  \node[main node, draw=none] (v2) [below of =v1]{};    
  \node[main node, draw=none] (v3) [right of =v2]{};    
  \node[main node] (6) [right of=v3] {$b$};
  \node[main node] (7) [below of=v2] {$c$};
  \node[main node] (8) [below of=7] {$d$};
  \node[main node] (10) [right of=5] {$h$};  
  \node[main node] (11) [right of=10] {$i$};  
  \node[main node] (12) [right of=8] {$f$};
  \node[main node] (13) [right of=12] {$g$};
  \node[main node] (14) [right of=13] {$j$};
  \path[every node/.style={font=\sffamily\small}]
    (6) edge node [] {} (12)     
    (6) edge node [] {} (13)     
    (7) edge node [] {} (5)     
    (12) edge node [] {} (5)     
    (7) edge node [] {} (12)     
    (7) edge node [] {} (6)     
    (12) edge node [] {} (13)         
    (5) edge node [] {} (10)         
    (11) edge node [] {} (10)         
    (11) edge node [] {} (14)         
    (13) edge node [] {} (14)         
    (5) edge node [] {} (6)     
    (12) edge node [] {} (8)     ;    
    \begin{pgfonlayer}{background}
        % Draw background box.
        \node[surround] (background) [fit = (v1) (14)] {};
        % Draw lines.
    \end{pgfonlayer}       
\end{tikzpicture}
\caption{Illustration of variable elimination. The initial undirected graphical model is displayed in the leftmost figure. Eliminating node $c$ in the reduces the number of tensors, as displayed in the middle figure. However, eliminating node $e$ introduces a tensor of rank $4$, as shown in the rightmost figure.}\label{fig:order}
\end{figure}
\end{center}

\section{Numerical Results}\label{numericalresults}

\begin{figure}[htbp]
\begin{tabular}{C{.25\textwidth}C{.25\textwidth}C{.25\textwidth}C{.25\textwidth}}
%%%%%%%%%%%%%%%%%%%%%% 1 %%%%%%%%%%%%%%%%%%%%%%%%%%%%%%%%%%%
\subfigure [Configuration 1] {
    	\resizebox{0.2\textwidth}{!}{%
    	\begin{tikzpicture}[node distance=2.0cm] %,scale=.5]
   	\scriptsize
	\foreach \i in {1,...,56}
	{
        		\pgfmathtruncatemacro{\y}{(\i - 1) / 7};
        		\pgfmathtruncatemacro{\x}{\i - 7 * \y};
        		\pgfmathtruncatemacro{\label}{\x + 7 * (7 - \y)};
        		\node[circle,draw=black,fill=white!80!black,minimum size=0.1]
        		(\label) at (1.5*\x,1.5*\y) {};
	}
% These draw commands are working as intended.
	\draw[ultra thick] (3) -- (4);
	\draw[ultra thick] (8) -- (9);
	\draw[ultra thick] (12) -- (13);
	\draw[ultra thick] (17) -- (18);
	\draw[ultra thick] (22) -- (23);
	\draw[ultra thick] (26) -- (27);
	\draw[ultra thick] (31) -- (32);
	\draw[ultra thick] (36) -- (37);
	\draw[ultra thick] (40) -- (41);
	\draw[ultra thick] (45) -- (46);
	\draw[ultra thick] (50) -- (51);
	\draw[ultra thick] (54) -- (55);
	\end{tikzpicture}
    	}
} & 
%%%%%%%%%%%%%%%%%%%%%% 2 %%%%%%%%%%%%%%%%%%%%%%%%%%%%%%%%%%%
\subfigure [Configuration 2] {
    	\resizebox{0.2\textwidth}{!}{%
    	\begin{tikzpicture}[node distance=2.0cm] %,scale=.5]
   	\scriptsize
	\foreach \i in {1,...,56}
	{
        		\pgfmathtruncatemacro{\y}{(\i - 1) / 7};
        		\pgfmathtruncatemacro{\x}{\i - 7 * \y};
        		\pgfmathtruncatemacro{\label}{\x + 7 * (7 - \y)};
        		\node[circle,draw=black,fill=white!80!black,minimum size=0.1]
        		(\label) at (1.5*\x,1.5*\y) {};
	}
% These draw commands are working as intended.
	\draw[ultra thick] (1) -- (2);
	\draw[ultra thick] (5) -- (6);
	\draw[ultra thick] (10) -- (11);
	\draw[ultra thick] (15) -- (16);
	\draw[ultra thick] (19) -- (20);
	\draw[ultra thick] (24) -- (25);
	\draw[ultra thick] (29) -- (30);
	\draw[ultra thick] (33) -- (34);
	\draw[ultra thick] (38) -- (39);
	\draw[ultra thick] (43) -- (44);
	\draw[ultra thick] (47) -- (48);
	\draw[ultra thick] (52) -- (53);
	\end{tikzpicture}
    	}
} &

%%%%%%%%%%%%%%%%%%%%%% 3 %%%%%%%%%%%%%%%%%%%%%%%%%%%%%%%%%%%
\subfigure [Configuration 3] {
  \resizebox{0.2\textwidth}{!}{%
    	\begin{tikzpicture}[node distance=2.0cm] %,scale=.5]
   	\scriptsize
	\foreach \i in {1,...,56}
	{
        		\pgfmathtruncatemacro{\y}{(\i - 1) / 7};
        		\pgfmathtruncatemacro{\x}{\i - 7 * \y};
        		\pgfmathtruncatemacro{\label}{\x + 7 * (7 - \y)};
        		\node[circle,draw=black,fill=white!80!black,minimum size=0.1]
        		(\label) at (1.5*\x,1.5*\y) {};
	}
% These draw commands are working as intended.
	\draw[ultra thick] (9) -- (16);
	\draw[ultra thick] (11) -- (18);
	\draw[ultra thick] (13) -- (20);
	\draw[ultra thick] (22) -- (29);
	\draw[ultra thick] (24) -- (31);
	\draw[ultra thick] (26) -- (33);
	\draw[ultra thick] (28) -- (35);
	\draw[ultra thick] (37) -- (44);
	\draw[ultra thick] (39) -- (46);
	\draw[ultra thick] (41) -- (48);
	\end{tikzpicture}
    	}
} & 

%%%%%%%%%%%%%%%%%%%%%% 4 %%%%%%%%%%%%%%%%%%%%%%%%%%%%%%%%%%%
\subfigure [Configuration 4] {
  \resizebox{0.2\textwidth}{!}{%
    	\begin{tikzpicture}[node distance=2.0cm] %,scale=.5]
   	\scriptsize
	\foreach \i in {1,...,56}
	{
        		\pgfmathtruncatemacro{\y}{(\i - 1) / 7};
        		\pgfmathtruncatemacro{\x}{\i - 7 * \y};
        		\pgfmathtruncatemacro{\label}{\x + 7 * (7 - \y)};
        		\node[circle,draw=black,fill=white!80!black,minimum size=0.1]
        		(\label) at (1.5*\x,1.5*\y) {};
	}
% These draw commands are working as intended.
	\draw[ultra thick] (8) -- (15);
	\draw[ultra thick] (10) -- (17);
	\draw[ultra thick] (12) -- (19);
	\draw[ultra thick] (14) -- (21);
	\draw[ultra thick] (23) -- (30);
	\draw[ultra thick] (25) -- (32);
	\draw[ultra thick] (27) -- (34);
	\draw[ultra thick] (36) -- (43);
	\draw[ultra thick] (38) -- (45);
	\draw[ultra thick] (40) -- (47);
	\draw[ultra thick] (42) -- (49);
	\end{tikzpicture}
    	}
} \\
\subfigure [Configuration 5] {
    	\resizebox{0.2\textwidth}{!}{%
    	\begin{tikzpicture}[node distance=2.0cm] %,scale=.5]
   	\scriptsize
	\foreach \i in {1,...,56}
	{
        		\pgfmathtruncatemacro{\y}{(\i - 1) / 7};
        		\pgfmathtruncatemacro{\x}{\i - 7 * \y};
        		\pgfmathtruncatemacro{\label}{\x + 7 * (7 - \y)};
        		\node[circle,draw=black,fill=white!80!black,minimum size=0.1]
        		(\label) at (1.5*\x,1.5*\y) {};
	}
% These draw commands are working as intended.
	\draw[ultra thick] (4) -- (5);
	\draw[ultra thick] (9) -- (10);
	\draw[ultra thick] (13) -- (14);
	\draw[ultra thick] (18) -- (19);
	\draw[ultra thick] (23) -- (24);
	\draw[ultra thick] (27) -- (28);
	\draw[ultra thick] (32) -- (33);
	\draw[ultra thick] (37) -- (38);
	\draw[ultra thick] (41) -- (42);
	\draw[ultra thick] (46) -- (47);
	\draw[ultra thick] (51) -- (52);
	\draw[ultra thick] (55) -- (56);
	\end{tikzpicture}
    	}
} & 
%%%%%%%%%%%%%%%%%%%%%% 2 %%%%%%%%%%%%%%%%%%%%%%%%%%%%%%%%%%%
\subfigure [Configuration 6] {
    	\resizebox{0.2\textwidth}{!}{%
    	\begin{tikzpicture}[node distance=2.0cm] %,scale=.5]
   	\scriptsize
	\foreach \i in {1,...,56}
	{
        		\pgfmathtruncatemacro{\y}{(\i - 1) / 7};
        		\pgfmathtruncatemacro{\x}{\i - 7 * \y};
        		\pgfmathtruncatemacro{\label}{\x + 7 * (7 - \y)};
        		\node[circle,draw=black,fill=white!80!black,minimum size=0.1]
        		(\label) at (1.5*\x,1.5*\y) {};
	}
% These draw commands are working as intended.
	\draw[ultra thick] (2) -- (3);
	\draw[ultra thick] (6) -- (7);
	\draw[ultra thick] (11) -- (12);
	\draw[ultra thick] (16) -- (17);
	\draw[ultra thick] (20) -- (21);
	\draw[ultra thick] (25) -- (26);
	\draw[ultra thick] (30) -- (31);
	\draw[ultra thick] (34) -- (35);
	\draw[ultra thick] (39) -- (40);
	\draw[ultra thick] (44) -- (45);
	\draw[ultra thick] (48) -- (49);
	\draw[ultra thick] (53) -- (54);
	\end{tikzpicture}
    	}
} &

%%%%%%%%%%%%%%%%%%%%%% 3 %%%%%%%%%%%%%%%%%%%%%%%%%%%%%%%%%%%
\subfigure [Configuration 7] {
  \resizebox{0.2\textwidth}{!}{%
    	\begin{tikzpicture}[node distance=2.0cm] %,scale=.5]
   	\scriptsize
	\foreach \i in {1,...,56}
	{
        		\pgfmathtruncatemacro{\y}{(\i - 1) / 7};
        		\pgfmathtruncatemacro{\x}{\i - 7 * \y};
        		\pgfmathtruncatemacro{\label}{\x + 7 * (7 - \y)};
        		\node[circle,draw=black,fill=white!80!black,minimum size=0.1]
        		(\label) at (1.5*\x,1.5*\y) {};
	}
% These draw commands are working as intended.
	\draw[ultra thick] (1) -- (8);
	\draw[ultra thick] (3) -- (10);
	\draw[ultra thick] (5) -- (12);
	\draw[ultra thick] (7) -- (14);
	\draw[ultra thick] (16) -- (23);
	\draw[ultra thick] (18) -- (25);
	\draw[ultra thick] (20) -- (27);
	\draw[ultra thick] (29) -- (36);
	\draw[ultra thick] (31) -- (38);
	\draw[ultra thick] (33) -- (40);
	\draw[ultra thick] (35) -- (42);
	\draw[ultra thick] (44) -- (51);
	\draw[ultra thick] (46) -- (53);
	\draw[ultra thick] (48) -- (55);
	\end{tikzpicture}
    	}
} & 

%%%%%%%%%%%%%%%%%%%%%% 4 %%%%%%%%%%%%%%%%%%%%%%%%%%%%%%%%%%%
\subfigure [Configuration 8] {
  \resizebox{0.2\textwidth}{!}{%
    	\begin{tikzpicture}[node distance=2.0cm] %,scale=.5]
   	\scriptsize
	\foreach \i in {1,...,56}
	{
        		\pgfmathtruncatemacro{\y}{(\i - 1) / 7};
        		\pgfmathtruncatemacro{\x}{\i - 7 * \y};
        		\pgfmathtruncatemacro{\label}{\x + 7 * (7 - \y)};
        		\node[circle,draw=black,fill=white!80!black,minimum size=0.1]
        		(\label) at (1.5*\x,1.5*\y) {};
	}
% These draw commands are working as intended.
	\draw[ultra thick] (2) -- (9);
	\draw[ultra thick] (4) -- (11);
	\draw[ultra thick] (6) -- (13);
	\draw[ultra thick] (15) -- (22);
	\draw[ultra thick] (17) -- (24);
	\draw[ultra thick] (19) -- (26);
	\draw[ultra thick] (21) -- (28);
	\draw[ultra thick] (30) -- (37);
	\draw[ultra thick] (32) -- (39);
	\draw[ultra thick] (34) -- (41);
	\draw[ultra thick] (43) -- (50);
	\draw[ultra thick] (45) -- (52);
	\draw[ultra thick] (47) -- (54);
	\draw[ultra thick] (49) -- (56);
\	\end{tikzpicture}
    	}
} \\
\end{tabular}
\caption{Layouts of CZ gates in the $8$ configurations producing supremacy circuits operating on an $8\times 7$ lattice of $56$ qubits \cite{BIS+16}.}
\label{fig:supremacy}
\end{figure}
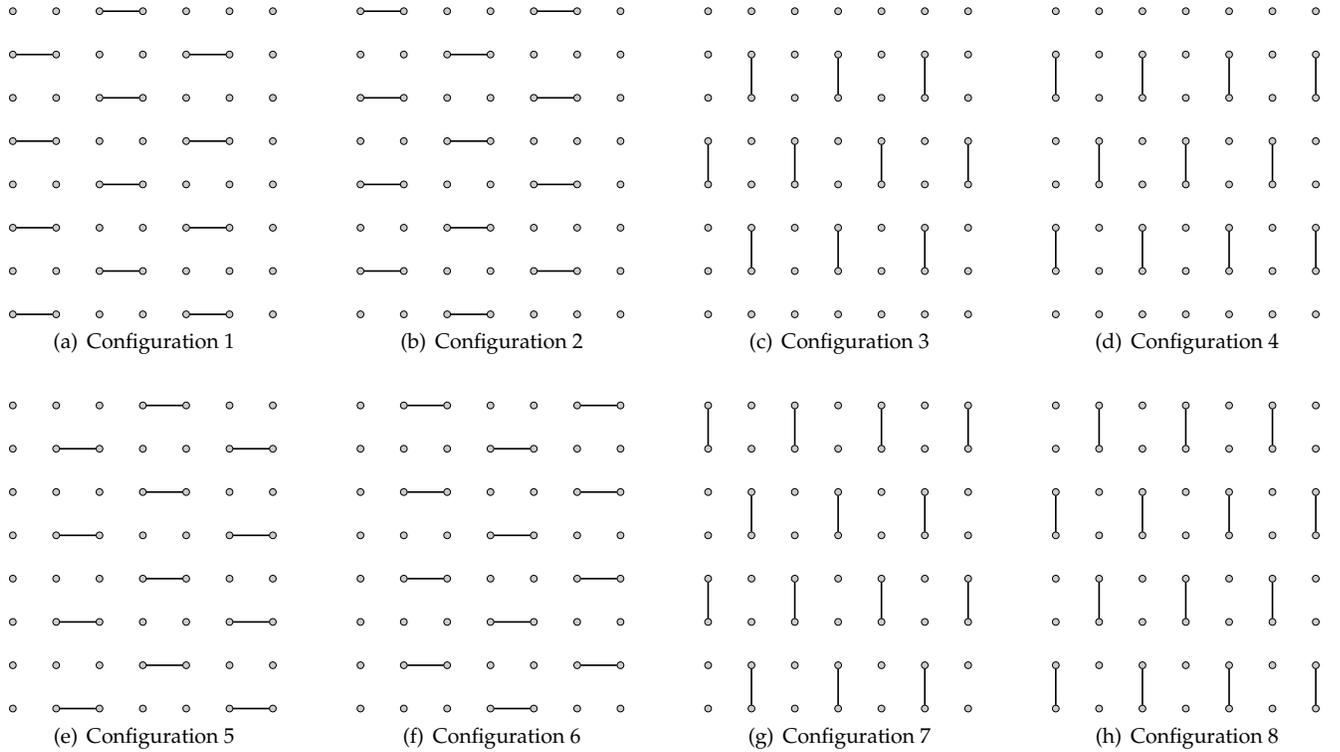

Although our algorithm can be applied to any quantum circuit, we will use the circuit family proposed for quantum supremacy in \cite{BIS+16} as test cases. These circuits were chosen through numerical optimizations to minimize the convergence time to the Porter-Thomas distribution. For the sake of completeness, we next recall the rules used to construct these quantum supremacy circuits.
\begin{enumerate}[\hspace{2cm}]\label{rules}
\item [(Rule 1)] Begin with a cycle of all Hadamard gates.
\item [(Rule 2)] Place CZ gates alternating between the 8 configurations shown in Figure~\ref{fig:supremacy}.
\item [(Rule 3)] Place single-qubit gates chosen at random from the set $\{\sqrt{\mathrm{X}}, \sqrt{\mathrm{Y}}, \mathrm{T}\}$ at the remaining positions, conditioned on the following restrictions. 
\begin{enumerate}
\item [(3.A)] Place a gate at a qubit only if that qubit is occupied by a CZ gate in the previous cycle.
\item [(3.B)] Place a $\mathrm{T}$-gate at a qubit if there are no single-qubit gates in the previous cycles at that location, except for the initial cycle of Hadamard gates.
\end{enumerate}
\end{enumerate}

We ran our tests on a small portion of the cluster provided by the Data Infrastructure and Search Technology Division of the Alibaba Group. It consists of 10,000 machines, each with 96 Intel Skylake Xeon Platinum 8163 vCPUs @ 2.5 GHz and with $512$ GB of memory. We requested $131072$ nodes via Docker containers and $8$ GB of memory for each node. We divided the circuit simulation task into $131072=2^{17}$ subtasks as described in Subsection~\ref{subsec:algorithm}. If the circuit is not that large, or more specifically, if the undirected graph has treewidth less than $28$, then each node has sufficient memory for the subtask. For larger circuits, if these subtasks require more than $8$ GB of memory, we further divide each subtask into sub-subtasks until the memory requirement can be met by each node. In this case, we run multiple sub-subtasks sequentially on each node and count the total running time of these sequential sub-subtasks on each node. 

We generate $1000$ random circuits operating on $n\times n$ lattices of depth $d$ according to the rules described above.  We test each circuit on the $131072$-node cluster and calculate the running time if it succeeds.  If it succeeds for $80\%$ of these circuits, we count the $80\%$ percentile running time for these instances, and claim that a typical circuit of size $n\times n\times d$ can be classically simulated within that time on our cluster \cite{BKM+14}. We use double precision since Python's built-in \texttt{float} and \texttt{complex} types have double precision.

\begin{figure}[htb!]
\centering
\includegraphics[width = 0.6\textwidth]{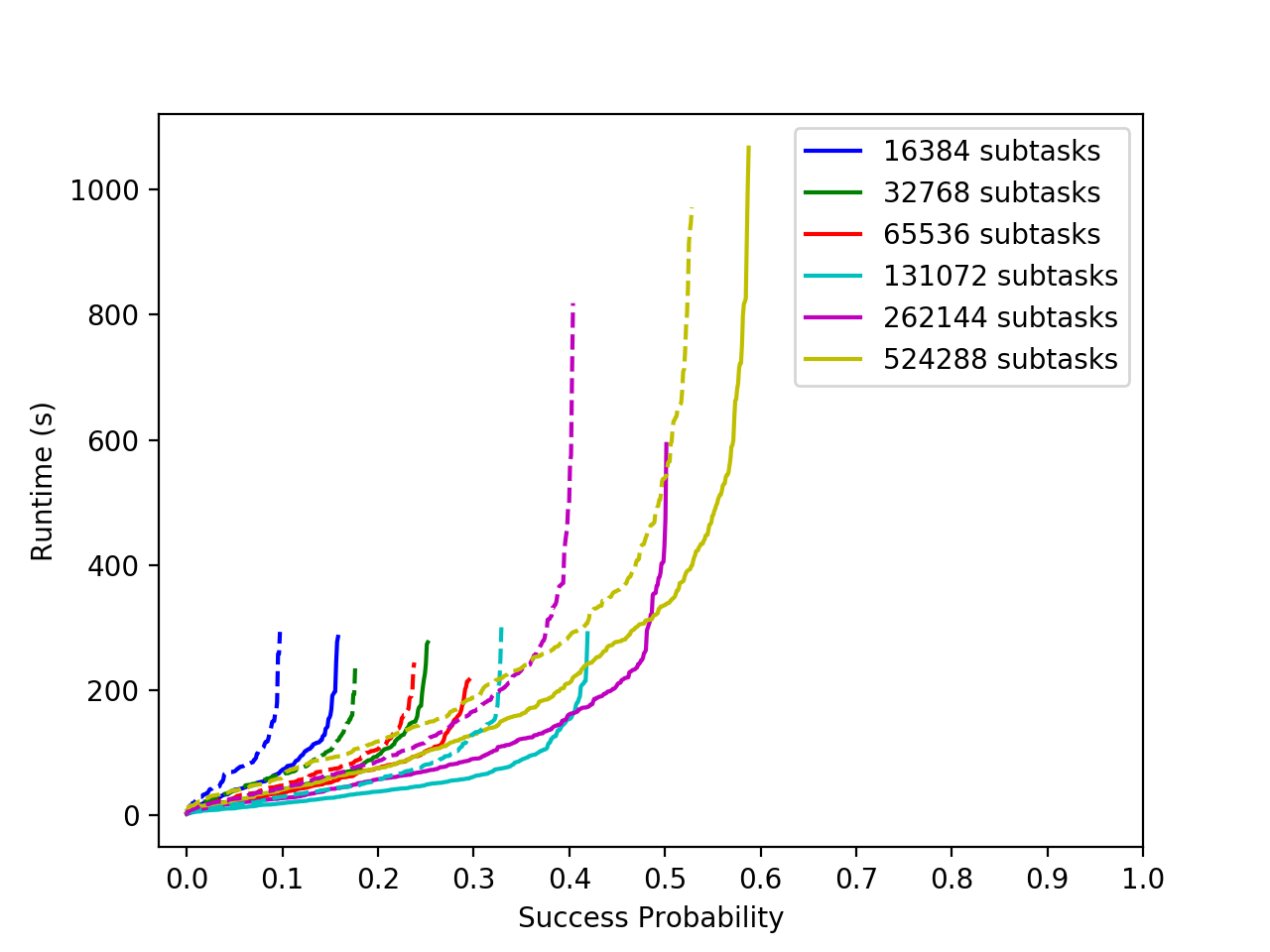}
\caption{The runtime versus success probability plot for our simulator acting on a $10 \times 10$ grid of depths $34$ and $35$.  Lines with different colors represent different numbers of subtasks that we divide our simulation task into. Regular lines and dashed lines represent depths $34$ and $35$, respectively. More subtasks always help to increase the success probability, until we reach the limit of our processors. After that, in order to improve the success probability, we have to tolerate a longer running time. }
\label{The }
\end{figure}

\section{Application to Random Circuit Sampling}\label{nosupremacy}

We next apply our simulator in the context of quantum supremacy from random circuit sampling.  We argue that, even for idealized gate fidelities on superconducting circuits, any circuit with a reasonable output fidelity may be simulated classically.

We study circuit fidelity by using the digital error model where each quantum gate is followed by an error channel \cite{BLK+15}.  The circuit fidelity for this model can be coarsely estimated as 
\begin{eqnarray*}
\alpha \approx \exp(-\epsilon_1 \times g_1 -\epsilon_2 \times g_2-\epsilon_{init} \times N -\epsilon_{mes} \times N)
\end{eqnarray*}
where $\epsilon_1, \epsilon_2 \ll 1$ are the Pauli error rates for single- and two-qubit gates, $\epsilon_{init}, \epsilon_{mes} \ll 1$ are the initialization and measurement error rates, $g_1, g_2$ are number of single- and two-qubit gates in the circuit, and $N$ is the number of qubits. In \cite{BIS+16}, the same technique was adopted to estimate the largest depth of circuits with $7\times 7$ qubits that can be demonstrated in state-of-art superconducting quantum computers.

\begin{figure}[htb!] % not h only
\centering
\includegraphics[width = 0.8\textwidth]{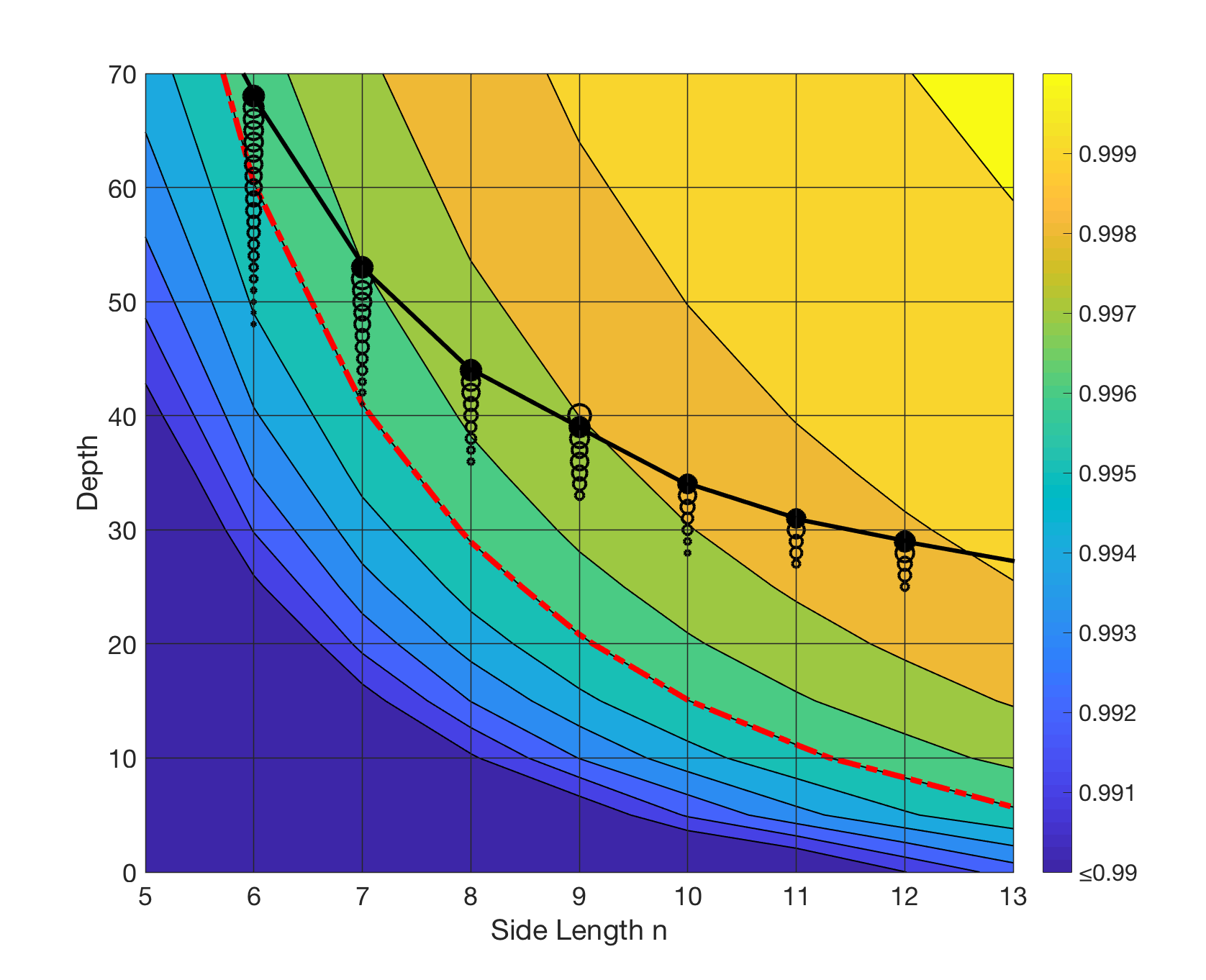}
\caption{ A scatter plot of the runtime superimposed on a contour graph of the two-qubit gate fidelities required for at least a $5\%$ circuit fidelity.  The $y$-axis represents the depth while the $x$-axis represents the side length of a square circuit, so that the total number of qubits being simulated is $n^2$.  The color scale specifies the two-qubit gate fidelity; for example, a $99\%$ two-qubit gate fidelity can only achieve a $5\%$ circuit fidelity in the bottom-left-most blue region, a $99.1\%$ two-qubit gate fidelity can only achieve a $5\%$ circuit fidelity up through the first band, and so on.  The dashed red curve represents the maximum circuit size achievable with $99.5\%$ two-qubit gate fidelity and at least a $5\%$ circuit fidelity.  \\\hspace{\textwidth} Each black circle represents the $80$th percentile runtime for that circuit size and depth over $1000$ samples, with the size of the circle representing the value of the runtime.  The circle size is plotted on a log-scale. For reference, the smallest circles represent approximately one second of runtime and the filled circles represent at most two hours of runtime.  The largest circle at $9 \times 9 \times 40$ represents approximately $13$ hours of runtime.}
\label{finalgraph}
\end{figure}

For simplicity, we make the physically-motivated assumption that $\epsilon_2=\epsilon_{init}=\epsilon_{mes}=\epsilon$ and $\epsilon_1=\frac{\epsilon}{10}$. For an $m\times n \times d$ circuit, the number of two-qubit gates in the configurations specified by Figure~\ref{fig:supremacy} can be estimated as $$g_2 \approx \frac{d}{8}\left((m-1)n + m(n-1)\right),$$ and this is precise whenever $d|8$.  Furthermore, we can approximate $$g_1 \approx \frac{d}{8}\left(3mn - (m+n +1)\right)-\frac14mn$$ when our circuit is drawn from that same family defined in \cite{BIS+16}.  Then, assuming $m \approx n$ and taking into account the initial application of Hadamard gates along with the first layer, we can approximate $$\alpha \approx \exp\left(-\left[\frac{d}{4}(N - \sqrt{N}) + 2N + \frac{d}{80}(3N - 2\sqrt{N} - 1) + \frac{3}{40}N\right]\epsilon\right).$$
State-of-the-art superconducting technology is expected to achieve two-qubit gate fidelities of at most $99.5 \%$  \cite{barends2014logic,kelly2015state}, corresponding to $\epsilon = 0.005$.  We plot our simulator bounds, given less than two hours of runtime, superimposed on the two-qubit gate fidelity graph required for a $5\%$ circuit fidelity.  We observe that we are well above the supremacy threshold for a $99.5\%$ two-qubit gate fidelity, drawn as a red dashed line. These results are summarized in Figure~\ref{finalgraph}.

We re-emphasize that we are reporting runtimes for single-amplitude simulation.  In order to generate a full distribution, significant overhead may be required in generating many amplitudes and sufficient trials.  However, these runtimes give evidence that, subject to tolerating this overhead, quantum supremacy will be difficult to achieve within this particular framework.  Generating a large slice of amplitudes, similar to \cite{PGN+17}, would be an interesting future direction that could provide more robustness to this simulation model.

\section{Discussion}\label{discussion}
In summary, we have described a cluster-based algorithm for quantum circuit simulation. By appropriately choosing vertices to eliminate in the undirected graphical model, we can reduce the treewidth significantly compared to selecting vertices at random.  By running our algorithm on the cluster provided by the Data Infrastructure and Search Technology Division of the Alibaba Group, we simulated quantum supremacy circuits of size $6\times 6\times 68$, $7\times 7 \times 53$, $8\times 8 \times 44$, $9\times 9\times 40$, $10\times 10\times 35$, $11\times 11\times 31$, and $12\times 12 \times 27$. We did this using $131072$ processors and $1$ petabyte of memory.  Finally, we gave evidence that noisy random circuits for realistic physical parameters and circuit fidelities may be simulated classically.  This suggests that new approaches or even error-correction may be necessary in demonstrating quantum supremacy \cite{preskill2012quantum,BIS+16}.  Fortunately, towards this end, there are already alternative proposals \cite{neill2018blueprint,farhi2016quantum,bremner2017achieving}.

In terms of the further improving the algorithm, we plan to explore an alternative to the greedy algorithm used to optimize the vertex elimination ordering. We observe that the vertices chosen for elimination via the greedy algorithm are always connected by at least two edges corresponding to CZ gates. Although this partially explains why it is more efficient than simply eliminating a random vertex (since by this choice, two rank-$2$ tensors will be replaced by a single rank-$2$ tensor as $\mathrm{CZ}_{12}\times \mathrm{CZ}_{13}=\ket{0}\bra{0}_1\otimes \mathrm{I}_{23}+\ket{1}\bra{1}_1\otimes \mathrm{Z}_{2}\mathrm{Z}_{3}$), it is worth further exploring this choice.  More generally, one could also consider an algorithm that is optimized for a particular restricted circuit class.

We should also mention that there are several limitations to our current implementation, which may be improved in future work.
\begin{enumerate}
\item The \texttt{numpy} package cannot handle too many subscripts within the \texttt{einsum} function, which is due to \texttt{NPY\_MAXDIMS}, the maximum number of dimensions allowed in arrays. This is defined as $32$ in \texttt{numpy}. 
\item If we need to divide the circuit simulation task into many more subtasks than the number of processes we have access to, our naive algorithm will run many of these subtasks sequentially on each processor. This reduces the gains obtained through parallelization.
\item For larger circuits, it is very slow to use QuickBB to generate a good elimination ordering. This is not surprising: choosing an optimal ordering is an NP-hard problem. However, it is worth exploring other avenues towards generating good orderings according to our cost function. Currently, we only use QuickBB twice. When running QuickBB in between the removal of each vertex, we can simulate much larger circuits (e.g. of size $7\times 7 \times 60$), but with an appreciably longer running time. Potentially, if we could find a faster algorithm, then we could run that algorithm between the removal of vertices. This may further simplify the subtasks at low cost. 
\end{enumerate}

One final consideration is our measure for characterizing the quality of our quantum devices.  Both here and in \cite{BIS+16}, we consider the circuit fidelity as a good measure of this quality.  However, there may be circuits with poor fidelity, but which still display quantum power with respect to a different metric. 

We hope that the increasing power of classical simulators pushes quantum processors to realize their near-term physical limits.  In the race to simulate quantum systems, we expect that our classical simulators will ultimately lose; this loss will be a demonstration of the intrinsic potential of quantum processing.

%-----------------------------------------------------------------------------%
\section*{Acknowledgements}
J. C. and F. Z. contributed equally to this project.  J. C.  would like to thank Mingcheng Chen for discussing an alternative approach to circuit simulation and the boundary of state-of-the-art superconducting hardware. He also thanks Mario Szegedy and Hua Xu for helpful conversations.  We thank our colleagues at Alibaba: Xiaowei Jiang, Xiaojun Jin, Guodong Yang, Min Zhang, Chi Zhang and Liang Ye for support with the computing facilities.  Finally, the authors thank Sergio Boixo, Igor Markov, and Mingsheng Ying for their valuable comments on an earlier draft of the paper.
%-----------------------------------------------------------------------------%

\bibliographystyle{apsrev4-1}
\bibliography{circuit}

\begin{thebibliography}{23}%
\makeatletter
\providecommand \@ifxundefined [1]{%
 \@ifx{#1\undefined}
}%
\providecommand \@ifnum [1]{%
 \ifnum #1\expandafter \@firstoftwo
 \else \expandafter \@secondoftwo
 \fi
}%
\providecommand \@ifx [1]{%
 \ifx #1\expandafter \@firstoftwo
 \else \expandafter \@secondoftwo
 \fi
}%
\providecommand \natexlab [1]{#1}%
\providecommand \enquote  [1]{``#1''}%
\providecommand \bibnamefont  [1]{#1}%
\providecommand \bibfnamefont [1]{#1}%
\providecommand \citenamefont [1]{#1}%
\providecommand \href@noop [0]{\@secondoftwo}%
\providecommand \href [0]{\begingroup \@sanitize@url \@href}%
\providecommand \@href[1]{\@@startlink{#1}\@@href}%
\providecommand \@@href[1]{\endgroup#1\@@endlink}%
\providecommand \@sanitize@url [0]{\catcode `\\12\catcode `\$12\catcode
  `\&12\catcode `\#12\catcode `\^12\catcode `\_12\catcode `\%12\relax}%
\providecommand \@@startlink[1]{}%
\providecommand \@@endlink[0]{}%
\providecommand \url  [0]{\begingroup\@sanitize@url \@url }%
\providecommand \@url [1]{\endgroup\@href {#1}{\urlprefix }}%
\providecommand \urlprefix  [0]{URL }%
\providecommand \Eprint [0]{\href }%
\providecommand \doibase [0]{http://dx.doi.org/}%
\providecommand \selectlanguage [0]{\@gobble}%
\providecommand \bibinfo  [0]{\@secondoftwo}%
\providecommand \bibfield  [0]{\@secondoftwo}%
\providecommand \translation [1]{[#1]}%
\providecommand \BibitemOpen [0]{}%
\providecommand \bibitemStop [0]{}%
\providecommand \bibitemNoStop [0]{.\EOS\space}%
\providecommand \EOS [0]{\spacefactor3000\relax}%
\providecommand \BibitemShut  [1]{\csname bibitem#1\endcsname}%
\let\auto@bib@innerbib\@empty
%</preamble>
\bibitem [{\citenamefont {Feynman}(1982)}]{feynman1982simulating}%
  \BibitemOpen
  \bibfield  {author} {\bibinfo {author} {\bibfnamefont {R.~P.}\ \bibnamefont
  {Feynman}},\ }\href@noop {} {\bibfield  {journal} {\bibinfo  {journal}
  {International Journal of Theoretical Physics}\ }\textbf {\bibinfo {volume}
  {21}},\ \bibinfo {pages} {467} (\bibinfo {year} {1982})}\BibitemShut
  {NoStop}%
\bibitem [{\citenamefont {Kelly}\ \emph {et~al.}(2015)\citenamefont {Kelly},
  \citenamefont {Barends}, \citenamefont {Fowler}, \citenamefont {Megrant},
  \citenamefont {Jeffrey}, \citenamefont {White}, \citenamefont {Sank},
  \citenamefont {Mutus}, \citenamefont {Campbell}, \citenamefont {Chen} \emph
  {et~al.}}]{kelly2015state}%
  \BibitemOpen
  \bibfield  {author} {\bibinfo {author} {\bibfnamefont {J.}~\bibnamefont
  {Kelly}}, \bibinfo {author} {\bibfnamefont {R.}~\bibnamefont {Barends}},
  \bibinfo {author} {\bibfnamefont {A.}~\bibnamefont {Fowler}}, \bibinfo
  {author} {\bibfnamefont {A.}~\bibnamefont {Megrant}}, \bibinfo {author}
  {\bibfnamefont {E.}~\bibnamefont {Jeffrey}}, \bibinfo {author} {\bibfnamefont
  {T.}~\bibnamefont {White}}, \bibinfo {author} {\bibfnamefont
  {D.}~\bibnamefont {Sank}}, \bibinfo {author} {\bibfnamefont {J.}~\bibnamefont
  {Mutus}}, \bibinfo {author} {\bibfnamefont {B.}~\bibnamefont {Campbell}},
  \bibinfo {author} {\bibfnamefont {Y.}~\bibnamefont {Chen}},  \emph {et~al.},\
  }\href@noop {} {\bibfield  {journal} {\bibinfo  {journal} {Nature}\ }\textbf
  {\bibinfo {volume} {519}},\ \bibinfo {pages} {66} (\bibinfo {year}
  {2015})}\BibitemShut {NoStop}%
\bibitem [{\citenamefont {Song}\ \emph {et~al.}(2017)\citenamefont {Song},
  \citenamefont {Xu}, \citenamefont {Liu}, \citenamefont {Yang}, \citenamefont
  {Zheng}, \citenamefont {Deng}, \citenamefont {Xie}, \citenamefont {Huang},
  \citenamefont {Guo}, \citenamefont {Zhang} \emph {et~al.}}]{song201710}%
  \BibitemOpen
  \bibfield  {author} {\bibinfo {author} {\bibfnamefont {C.}~\bibnamefont
  {Song}}, \bibinfo {author} {\bibfnamefont {K.}~\bibnamefont {Xu}}, \bibinfo
  {author} {\bibfnamefont {W.}~\bibnamefont {Liu}}, \bibinfo {author}
  {\bibfnamefont {C.-P.}\ \bibnamefont {Yang}}, \bibinfo {author}
  {\bibfnamefont {S.-B.}\ \bibnamefont {Zheng}}, \bibinfo {author}
  {\bibfnamefont {H.}~\bibnamefont {Deng}}, \bibinfo {author} {\bibfnamefont
  {Q.}~\bibnamefont {Xie}}, \bibinfo {author} {\bibfnamefont {K.}~\bibnamefont
  {Huang}}, \bibinfo {author} {\bibfnamefont {Q.}~\bibnamefont {Guo}}, \bibinfo
  {author} {\bibfnamefont {L.}~\bibnamefont {Zhang}},  \emph {et~al.},\
  }\href@noop {} {\bibfield  {journal} {\bibinfo  {journal} {Physical Review
  Letters}\ }\textbf {\bibinfo {volume} {119}},\ \bibinfo {pages} {180511}
  (\bibinfo {year} {2017})}\BibitemShut {NoStop}%
\bibitem [{\citenamefont {{De Raedt}}\ \emph {et~al.}(2007)\citenamefont {{De
  Raedt}}, \citenamefont {{Michielsen}}, \citenamefont {{De Raedt}},
  \citenamefont {{Trieu}}, \citenamefont {{Arnold}}, \citenamefont {{Richter}},
  \citenamefont {{Lippert}}, \citenamefont {{Watanabe}},\ and\ \citenamefont
  {{Ito}}}]{RMR+06}%
  \BibitemOpen
  \bibfield  {author} {\bibinfo {author} {\bibfnamefont {K.}~\bibnamefont {{De
  Raedt}}}, \bibinfo {author} {\bibfnamefont {K.}~\bibnamefont {{Michielsen}}},
  \bibinfo {author} {\bibfnamefont {H.}~\bibnamefont {{De Raedt}}}, \bibinfo
  {author} {\bibfnamefont {B.}~\bibnamefont {{Trieu}}}, \bibinfo {author}
  {\bibfnamefont {G.}~\bibnamefont {{Arnold}}}, \bibinfo {author}
  {\bibfnamefont {M.}~\bibnamefont {{Richter}}}, \bibinfo {author}
  {\bibfnamefont {T.}~\bibnamefont {{Lippert}}}, \bibinfo {author}
  {\bibfnamefont {H.}~\bibnamefont {{Watanabe}}}, \ and\ \bibinfo {author}
  {\bibfnamefont {N.}~\bibnamefont {{Ito}}},\ }\href {\doibase
  10.1016/j.cpc.2006.08.007} {\bibfield  {journal} {\bibinfo  {journal}
  {{Computer Physics Communications}}\ }\textbf {\bibinfo {volume} {176}},\
  \bibinfo {pages} {121} (\bibinfo {year} {2007})},\ \Eprint
  {http://arxiv.org/abs/quant-ph/0608239} {quant-ph/0608239} \BibitemShut
  {NoStop}%
\bibitem [{\citenamefont {H{\"a}ner}\ and\ \citenamefont
  {Steiger}(2017)}]{HS17}%
  \BibitemOpen
  \bibfield  {author} {\bibinfo {author} {\bibfnamefont {T.}~\bibnamefont
  {H{\"a}ner}}\ and\ \bibinfo {author} {\bibfnamefont {D.~S.}\ \bibnamefont
  {Steiger}},\ }in\ \href@noop {} {\emph {\bibinfo {booktitle} {Proceedings of
  the International Conference for High Performance Computing, Networking,
  Storage and Analysis}}}\ (\bibinfo {organization} {ACM},\ \bibinfo {year}
  {2017})\ p.~\bibinfo {pages} {33}\BibitemShut {NoStop}%
\bibitem [{\citenamefont {{Smelyanskiy}}\ \emph {et~al.}(2016)\citenamefont
  {{Smelyanskiy}}, \citenamefont {{Sawaya}},\ and\ \citenamefont
  {{Aspuru-Guzik}}}]{SSA16}%
  \BibitemOpen
  \bibfield  {author} {\bibinfo {author} {\bibfnamefont {M.}~\bibnamefont
  {{Smelyanskiy}}}, \bibinfo {author} {\bibfnamefont {N.~P.~D.}\ \bibnamefont
  {{Sawaya}}}, \ and\ \bibinfo {author} {\bibfnamefont {A.}~\bibnamefont
  {{Aspuru-Guzik}}},\ }\href@noop {} {\bibfield  {journal} {\bibinfo  {journal}
  {ArXiv e-prints}\ } (\bibinfo {year} {2016})},\ \Eprint
  {http://arxiv.org/abs/1601.07195} {arXiv:1601.07195 [quant-ph]} \BibitemShut
  {NoStop}%
\bibitem [{\citenamefont {{Pednault}}\ \emph {et~al.}(2017)\citenamefont
  {{Pednault}}, \citenamefont {{Gunnels}}, \citenamefont {{Nannicini}},
  \citenamefont {{Horesh}}, \citenamefont {{Magerlein}}, \citenamefont
  {{Solomonik}},\ and\ \citenamefont {{Wisnieff}}}]{PGN+17}%
  \BibitemOpen
  \bibfield  {author} {\bibinfo {author} {\bibfnamefont {E.}~\bibnamefont
  {{Pednault}}}, \bibinfo {author} {\bibfnamefont {J.~A.}\ \bibnamefont
  {{Gunnels}}}, \bibinfo {author} {\bibfnamefont {G.}~\bibnamefont
  {{Nannicini}}}, \bibinfo {author} {\bibfnamefont {L.}~\bibnamefont
  {{Horesh}}}, \bibinfo {author} {\bibfnamefont {T.}~\bibnamefont
  {{Magerlein}}}, \bibinfo {author} {\bibfnamefont {E.}~\bibnamefont
  {{Solomonik}}}, \ and\ \bibinfo {author} {\bibfnamefont {R.}~\bibnamefont
  {{Wisnieff}}},\ }\href@noop {} {\bibfield  {journal} {\bibinfo  {journal}
  {ArXiv e-prints}\ } (\bibinfo {year} {2017})},\ \Eprint
  {http://arxiv.org/abs/1710.05867} {arXiv:1710.05867 [quant-ph]} \BibitemShut
  {NoStop}%
\bibitem [{\citenamefont {{Boixo}}\ \emph {et~al.}(2017)\citenamefont
  {{Boixo}}, \citenamefont {{Isakov}}, \citenamefont {{Smelyanskiy}},\ and\
  \citenamefont {{Neven}}}]{BIS+17}%
  \BibitemOpen
  \bibfield  {author} {\bibinfo {author} {\bibfnamefont {S.}~\bibnamefont
  {{Boixo}}}, \bibinfo {author} {\bibfnamefont {S.~V.}\ \bibnamefont
  {{Isakov}}}, \bibinfo {author} {\bibfnamefont {V.~N.}\ \bibnamefont
  {{Smelyanskiy}}}, \ and\ \bibinfo {author} {\bibfnamefont {H.}~\bibnamefont
  {{Neven}}},\ }\href@noop {} {\bibfield  {journal} {\bibinfo  {journal} {ArXiv
  e-prints}\ } (\bibinfo {year} {2017})},\ \Eprint
  {http://arxiv.org/abs/1712.05384} {arXiv:1712.05384 [quant-ph]} \BibitemShut
  {NoStop}%
\bibitem [{\citenamefont {{Chen}}\ \emph {et~al.}(2018)\citenamefont {{Chen}},
  \citenamefont {{Zhou}}, \citenamefont {{Xue}}, \citenamefont {{Yang}},
  \citenamefont {{Guo}},\ and\ \citenamefont {{Guo}}}]{CZX+18}%
  \BibitemOpen
  \bibfield  {author} {\bibinfo {author} {\bibfnamefont {Z.-Y.}\ \bibnamefont
  {{Chen}}}, \bibinfo {author} {\bibfnamefont {Q.}~\bibnamefont {{Zhou}}},
  \bibinfo {author} {\bibfnamefont {C.}~\bibnamefont {{Xue}}}, \bibinfo
  {author} {\bibfnamefont {X.}~\bibnamefont {{Yang}}}, \bibinfo {author}
  {\bibfnamefont {G.-C.}\ \bibnamefont {{Guo}}}, \ and\ \bibinfo {author}
  {\bibfnamefont {G.-P.}\ \bibnamefont {{Guo}}},\ }\href@noop {} {\bibfield
  {journal} {\bibinfo  {journal} {ArXiv e-prints}\ } (\bibinfo {year}
  {2018})},\ \Eprint {http://arxiv.org/abs/1802.06952} {arXiv:1802.06952
  [quant-ph]} \BibitemShut {NoStop}%
\bibitem [{\citenamefont {{Li}}\ \emph {et~al.}(2018)\citenamefont {{Li}},
  \citenamefont {{Wu}}, \citenamefont {{Ying}}, \citenamefont {{Sun}},\ and\
  \citenamefont {{Yang}}}]{LWY+18}%
  \BibitemOpen
  \bibfield  {author} {\bibinfo {author} {\bibfnamefont {R.}~\bibnamefont
  {{Li}}}, \bibinfo {author} {\bibfnamefont {B.}~\bibnamefont {{Wu}}}, \bibinfo
  {author} {\bibfnamefont {M.}~\bibnamefont {{Ying}}}, \bibinfo {author}
  {\bibfnamefont {X.}~\bibnamefont {{Sun}}}, \ and\ \bibinfo {author}
  {\bibfnamefont {G.}~\bibnamefont {{Yang}}},\ }\href@noop {} {\bibfield
  {journal} {\bibinfo  {journal} {ArXiv e-prints}\ } (\bibinfo {year}
  {2018})},\ \Eprint {http://arxiv.org/abs/1804.04797} {arXiv:1804.04797
  [quant-ph]} \BibitemShut {NoStop}%
\bibitem [{\citenamefont {Aaronson}\ and\ \citenamefont
  {Chen}(2017)}]{aaronson2017complexity}%
  \BibitemOpen
  \bibfield  {author} {\bibinfo {author} {\bibfnamefont {S.}~\bibnamefont
  {Aaronson}}\ and\ \bibinfo {author} {\bibfnamefont {L.}~\bibnamefont
  {Chen}},\ }in\ \href@noop {} {\emph {\bibinfo {booktitle} {Proceedings of the
  32nd Computational Complexity Conference}}}\ (\bibinfo {organization}
  {Schloss Dagstuhl--Leibniz-Zentrum fuer Informatik},\ \bibinfo {year}
  {2017})\ p.~\bibinfo {pages} {22}\BibitemShut {NoStop}%
\bibitem [{\citenamefont {{Preskill}}(2018)}]{P18}%
  \BibitemOpen
  \bibfield  {author} {\bibinfo {author} {\bibfnamefont {J.}~\bibnamefont
  {{Preskill}}},\ }\href@noop {} {\bibfield  {journal} {\bibinfo  {journal}
  {ArXiv e-prints}\ } (\bibinfo {year} {2018})},\ \Eprint
  {http://arxiv.org/abs/1801.00862} {arXiv:1801.00862 [quant-ph]} \BibitemShut
  {NoStop}%
\bibitem [{\citenamefont {Boixo}\ \emph {et~al.}(2018)\citenamefont {Boixo},
  \citenamefont {Isakov}, \citenamefont {Smelyanskiy}, \citenamefont {Babbush},
  \citenamefont {Ding}, \citenamefont {Jiang}, \citenamefont {Bremner},
  \citenamefont {Martinis},\ and\ \citenamefont {Neven}}]{BIS+16}%
  \BibitemOpen
  \bibfield  {author} {\bibinfo {author} {\bibfnamefont {S.}~\bibnamefont
  {Boixo}}, \bibinfo {author} {\bibfnamefont {S.~V.}\ \bibnamefont {Isakov}},
  \bibinfo {author} {\bibfnamefont {V.~N.}\ \bibnamefont {Smelyanskiy}},
  \bibinfo {author} {\bibfnamefont {R.}~\bibnamefont {Babbush}}, \bibinfo
  {author} {\bibfnamefont {N.}~\bibnamefont {Ding}}, \bibinfo {author}
  {\bibfnamefont {Z.}~\bibnamefont {Jiang}}, \bibinfo {author} {\bibfnamefont
  {M.~J.}\ \bibnamefont {Bremner}}, \bibinfo {author} {\bibfnamefont {J.~M.}\
  \bibnamefont {Martinis}}, \ and\ \bibinfo {author} {\bibfnamefont
  {H.}~\bibnamefont {Neven}},\ }\href@noop {} {\bibfield  {journal} {\bibinfo
  {journal} {Nature Physics}\ ,\ \bibinfo {pages} {1}} (\bibinfo {year}
  {2018})}\BibitemShut {NoStop}%
\bibitem [{\citenamefont {Markov}\ and\ \citenamefont {Shi}(2008)}]{MS08}%
  \BibitemOpen
  \bibfield  {author} {\bibinfo {author} {\bibfnamefont {I.~L.}\ \bibnamefont
  {Markov}}\ and\ \bibinfo {author} {\bibfnamefont {Y.}~\bibnamefont {Shi}},\
  }\href {\doibase 10.1137/050644756} {\bibfield  {journal} {\bibinfo
  {journal} {SIAM Journal on Computing}\ }\textbf {\bibinfo {volume} {38}},\
  \bibinfo {pages} {963} (\bibinfo {year} {2008})}\BibitemShut {NoStop}%
\bibitem [{\citenamefont {Li}\ \emph {et~al.}(2018)\citenamefont {Li},
  \citenamefont {Wu}, \citenamefont {Ying}, \citenamefont {Sun},\ and\
  \citenamefont {Yang}}]{li2018quantum}%
  \BibitemOpen
  \bibfield  {author} {\bibinfo {author} {\bibfnamefont {R.}~\bibnamefont
  {Li}}, \bibinfo {author} {\bibfnamefont {B.}~\bibnamefont {Wu}}, \bibinfo
  {author} {\bibfnamefont {M.}~\bibnamefont {Ying}}, \bibinfo {author}
  {\bibfnamefont {X.}~\bibnamefont {Sun}}, \ and\ \bibinfo {author}
  {\bibfnamefont {G.}~\bibnamefont {Yang}},\ }\href@noop {} {\bibfield
  {journal} {\bibinfo  {journal} {arXiv preprint arXiv:1804.04797}\ } (\bibinfo
  {year} {2018})}\BibitemShut {NoStop}%
\bibitem [{\citenamefont {Gogate}\ and\ \citenamefont {Dechter}(2004)}]{VR04}%
  \BibitemOpen
  \bibfield  {author} {\bibinfo {author} {\bibfnamefont {V.}~\bibnamefont
  {Gogate}}\ and\ \bibinfo {author} {\bibfnamefont {R.}~\bibnamefont
  {Dechter}},\ }in\ \href@noop {} {\emph {\bibinfo {booktitle} {UAI}}}\
  (\bibinfo {year} {2004})\ \Eprint {http://arxiv.org/abs/1207.4109}
  {arXiv:1207.4109} \BibitemShut {NoStop}%
\bibitem [{\citenamefont {{Barends}}\ \emph {et~al.}(2014)\citenamefont
  {{Barends}}, \citenamefont {{Kelly}}, \citenamefont {{Megrant}},
  \citenamefont {{Veitia}}, \citenamefont {{Sank}}, \citenamefont {{Jeffrey}},
  \citenamefont {{White}}, \citenamefont {{Mutus}}, \citenamefont {{Fowler}},
  \citenamefont {{Campbell}}, \citenamefont {{Chen}}, \citenamefont {{Chen}},
  \citenamefont {{Chiaro}}, \citenamefont {{Dunsworth}}, \citenamefont
  {{Neill}}, \citenamefont {{O'Malley}}, \citenamefont {{Roushan}},
  \citenamefont {{Vainsencher}}, \citenamefont {{Wenner}}, \citenamefont
  {{Korotkov}}, \citenamefont {{Cleland}},\ and\ \citenamefont
  {{Martinis}}}]{BKM+14}%
  \BibitemOpen
  \bibfield  {author} {\bibinfo {author} {\bibfnamefont {R.}~\bibnamefont
  {{Barends}}}, \bibinfo {author} {\bibfnamefont {J.}~\bibnamefont {{Kelly}}},
  \bibinfo {author} {\bibfnamefont {A.}~\bibnamefont {{Megrant}}}, \bibinfo
  {author} {\bibfnamefont {A.}~\bibnamefont {{Veitia}}}, \bibinfo {author}
  {\bibfnamefont {D.}~\bibnamefont {{Sank}}}, \bibinfo {author} {\bibfnamefont
  {E.}~\bibnamefont {{Jeffrey}}}, \bibinfo {author} {\bibfnamefont {T.~C.}\
  \bibnamefont {{White}}}, \bibinfo {author} {\bibfnamefont {J.}~\bibnamefont
  {{Mutus}}}, \bibinfo {author} {\bibfnamefont {A.~G.}\ \bibnamefont
  {{Fowler}}}, \bibinfo {author} {\bibfnamefont {B.}~\bibnamefont
  {{Campbell}}}, \bibinfo {author} {\bibfnamefont {Y.}~\bibnamefont {{Chen}}},
  \bibinfo {author} {\bibfnamefont {Z.}~\bibnamefont {{Chen}}}, \bibinfo
  {author} {\bibfnamefont {B.}~\bibnamefont {{Chiaro}}}, \bibinfo {author}
  {\bibfnamefont {A.}~\bibnamefont {{Dunsworth}}}, \bibinfo {author}
  {\bibfnamefont {C.}~\bibnamefont {{Neill}}}, \bibinfo {author} {\bibfnamefont
  {P.}~\bibnamefont {{O'Malley}}}, \bibinfo {author} {\bibfnamefont
  {P.}~\bibnamefont {{Roushan}}}, \bibinfo {author} {\bibfnamefont
  {A.}~\bibnamefont {{Vainsencher}}}, \bibinfo {author} {\bibfnamefont
  {J.}~\bibnamefont {{Wenner}}}, \bibinfo {author} {\bibfnamefont {A.~N.}\
  \bibnamefont {{Korotkov}}}, \bibinfo {author} {\bibfnamefont {A.~N.}\
  \bibnamefont {{Cleland}}}, \ and\ \bibinfo {author} {\bibfnamefont {J.~M.}\
  \bibnamefont {{Martinis}}},\ }\href {\doibase 10.1038/nature13171} {\bibfield
   {journal} {\bibinfo  {journal} {\nat}\ }\textbf {\bibinfo {volume} {508}},\
  \bibinfo {pages} {500} (\bibinfo {year} {2014})},\ \Eprint
  {http://arxiv.org/abs/1402.4848} {arXiv:1402.4848 [quant-ph]} \BibitemShut
  {NoStop}%
\bibitem [{\citenamefont {Barends}\ \emph {et~al.}(2015)\citenamefont
  {Barends}, \citenamefont {Lamata}, \citenamefont {Kelly}, \citenamefont
  {Garc{\'\i}a-{\'A}lvarez}, \citenamefont {Fowler}, \citenamefont {Megrant},
  \citenamefont {Jeffrey}, \citenamefont {White}, \citenamefont {Sank},
  \citenamefont {Mutus}, \citenamefont {Campbell}, \citenamefont {Chen},
  \citenamefont {Chen}, \citenamefont {Chiaro}, \citenamefont {Dunsworth},
  \citenamefont {Hoi}, \citenamefont {Neill}, \citenamefont {O'Malley},
  \citenamefont {Quintana}, \citenamefont {Roushan}, \citenamefont
  {Vainsencher}, \citenamefont {Wenner}, \citenamefont {Solano},\ and\
  \citenamefont {Martinis}}]{BLK+15}%
  \BibitemOpen
  \bibfield  {author} {\bibinfo {author} {\bibfnamefont {R.}~\bibnamefont
  {Barends}}, \bibinfo {author} {\bibfnamefont {L.}~\bibnamefont {Lamata}},
  \bibinfo {author} {\bibfnamefont {J.}~\bibnamefont {Kelly}}, \bibinfo
  {author} {\bibfnamefont {L.}~\bibnamefont {Garc{\'\i}a-{\'A}lvarez}},
  \bibinfo {author} {\bibfnamefont {A.~G.}\ \bibnamefont {Fowler}}, \bibinfo
  {author} {\bibfnamefont {A.}~\bibnamefont {Megrant}}, \bibinfo {author}
  {\bibfnamefont {E.}~\bibnamefont {Jeffrey}}, \bibinfo {author} {\bibfnamefont
  {T.~C.}\ \bibnamefont {White}}, \bibinfo {author} {\bibfnamefont
  {D.}~\bibnamefont {Sank}}, \bibinfo {author} {\bibfnamefont {J.~Y.}\
  \bibnamefont {Mutus}}, \bibinfo {author} {\bibfnamefont {B.}~\bibnamefont
  {Campbell}}, \bibinfo {author} {\bibfnamefont {Y.}~\bibnamefont {Chen}},
  \bibinfo {author} {\bibfnamefont {Z.}~\bibnamefont {Chen}}, \bibinfo {author}
  {\bibfnamefont {B.}~\bibnamefont {Chiaro}}, \bibinfo {author} {\bibfnamefont
  {A.}~\bibnamefont {Dunsworth}}, \bibinfo {author} {\bibfnamefont {I.~C.}\
  \bibnamefont {Hoi}}, \bibinfo {author} {\bibfnamefont {C.}~\bibnamefont
  {Neill}}, \bibinfo {author} {\bibfnamefont {P.~J.~J.}\ \bibnamefont
  {O'Malley}}, \bibinfo {author} {\bibfnamefont {C.}~\bibnamefont {Quintana}},
  \bibinfo {author} {\bibfnamefont {P.}~\bibnamefont {Roushan}}, \bibinfo
  {author} {\bibfnamefont {A.}~\bibnamefont {Vainsencher}}, \bibinfo {author}
  {\bibfnamefont {J.}~\bibnamefont {Wenner}}, \bibinfo {author} {\bibfnamefont
  {E.}~\bibnamefont {Solano}}, \ and\ \bibinfo {author} {\bibfnamefont {J.~M.}\
  \bibnamefont {Martinis}},\ }\href {http://dx.doi.org/10.1038/ncomms8654}
  {\bibfield  {journal} {\bibinfo  {journal} {Nature Communications}\ }\textbf
  {\bibinfo {volume} {6}},\ \bibinfo {pages} {7654 EP } (\bibinfo {year}
  {2015})}\BibitemShut {NoStop}%
\bibitem [{\citenamefont {Barends}\ \emph {et~al.}(2014)\citenamefont
  {Barends}, \citenamefont {Kelly}, \citenamefont {Megrant}, \citenamefont
  {Veitia}, \citenamefont {Sank}, \citenamefont {Jeffrey}, \citenamefont
  {White}, \citenamefont {Mutus}, \citenamefont {Fowler}, \citenamefont
  {Campbell} \emph {et~al.}}]{barends2014logic}%
  \BibitemOpen
  \bibfield  {author} {\bibinfo {author} {\bibfnamefont {R.}~\bibnamefont
  {Barends}}, \bibinfo {author} {\bibfnamefont {J.}~\bibnamefont {Kelly}},
  \bibinfo {author} {\bibfnamefont {A.}~\bibnamefont {Megrant}}, \bibinfo
  {author} {\bibfnamefont {A.}~\bibnamefont {Veitia}}, \bibinfo {author}
  {\bibfnamefont {D.}~\bibnamefont {Sank}}, \bibinfo {author} {\bibfnamefont
  {E.}~\bibnamefont {Jeffrey}}, \bibinfo {author} {\bibfnamefont
  {T.}~\bibnamefont {White}}, \bibinfo {author} {\bibfnamefont
  {J.}~\bibnamefont {Mutus}}, \bibinfo {author} {\bibfnamefont
  {A.}~\bibnamefont {Fowler}}, \bibinfo {author} {\bibfnamefont
  {B.}~\bibnamefont {Campbell}},  \emph {et~al.},\ }\href@noop {} {\bibfield
  {journal} {\bibinfo  {journal} {arXiv preprint arXiv:1402.4848}\ } (\bibinfo
  {year} {2014})}\BibitemShut {NoStop}%
\bibitem [{\citenamefont {Preskill}(2012)}]{preskill2012quantum}%
  \BibitemOpen
  \bibfield  {author} {\bibinfo {author} {\bibfnamefont {J.}~\bibnamefont
  {Preskill}},\ }\href@noop {} {\bibfield  {journal} {\bibinfo  {journal}
  {arXiv preprint arXiv:1203.5813}\ } (\bibinfo {year} {2012})}\BibitemShut
  {NoStop}%
\bibitem [{\citenamefont {Neill}\ \emph {et~al.}(2018)\citenamefont {Neill},
  \citenamefont {Roushan}, \citenamefont {Kechedzhi}, \citenamefont {Boixo},
  \citenamefont {Isakov}, \citenamefont {Smelyanskiy}, \citenamefont {Megrant},
  \citenamefont {Chiaro}, \citenamefont {Dunsworth}, \citenamefont {Arya} \emph
  {et~al.}}]{neill2018blueprint}%
  \BibitemOpen
  \bibfield  {author} {\bibinfo {author} {\bibfnamefont {C.}~\bibnamefont
  {Neill}}, \bibinfo {author} {\bibfnamefont {P.}~\bibnamefont {Roushan}},
  \bibinfo {author} {\bibfnamefont {K.}~\bibnamefont {Kechedzhi}}, \bibinfo
  {author} {\bibfnamefont {S.}~\bibnamefont {Boixo}}, \bibinfo {author}
  {\bibfnamefont {S.}~\bibnamefont {Isakov}}, \bibinfo {author} {\bibfnamefont
  {V.}~\bibnamefont {Smelyanskiy}}, \bibinfo {author} {\bibfnamefont
  {A.}~\bibnamefont {Megrant}}, \bibinfo {author} {\bibfnamefont
  {B.}~\bibnamefont {Chiaro}}, \bibinfo {author} {\bibfnamefont
  {A.}~\bibnamefont {Dunsworth}}, \bibinfo {author} {\bibfnamefont
  {K.}~\bibnamefont {Arya}},  \emph {et~al.},\ }\href@noop {} {\bibfield
  {journal} {\bibinfo  {journal} {Science}\ }\textbf {\bibinfo {volume}
  {360}},\ \bibinfo {pages} {195} (\bibinfo {year} {2018})}\BibitemShut
  {NoStop}%
\bibitem [{\citenamefont {Farhi}\ and\ \citenamefont
  {Harrow}(2016)}]{farhi2016quantum}%
  \BibitemOpen
  \bibfield  {author} {\bibinfo {author} {\bibfnamefont {E.}~\bibnamefont
  {Farhi}}\ and\ \bibinfo {author} {\bibfnamefont {A.~W.}\ \bibnamefont
  {Harrow}},\ }\href@noop {} {\bibfield  {journal} {\bibinfo  {journal} {arXiv
  preprint arXiv:1602.07674}\ } (\bibinfo {year} {2016})}\BibitemShut {NoStop}%
\bibitem [{\citenamefont {Bremner}\ \emph {et~al.}(2017)\citenamefont
  {Bremner}, \citenamefont {Montanaro},\ and\ \citenamefont
  {Shepherd}}]{bremner2017achieving}%
  \BibitemOpen
  \bibfield  {author} {\bibinfo {author} {\bibfnamefont {M.~J.}\ \bibnamefont
  {Bremner}}, \bibinfo {author} {\bibfnamefont {A.}~\bibnamefont {Montanaro}},
  \ and\ \bibinfo {author} {\bibfnamefont {D.~J.}\ \bibnamefont {Shepherd}},\
  }\href@noop {} {\bibfield  {journal} {\bibinfo  {journal} {Quantum}\ }\textbf
  {\bibinfo {volume} {1}},\ \bibinfo {pages} {8} (\bibinfo {year}
  {2017})}\BibitemShut {NoStop}%
\end{thebibliography}%

\appendix

%% Start-Of-Trailer

\end{document}